\documentstyle [12pt,eqsecnum,aps,amsfonts] {revtex}
\input epsf
\newcommand{\beq}{\begin{equation}}
\newcommand{\eeq}{\end{equation}}
\newcommand{\beqs}{\begin{eqnarray}}
\newcommand{\eeqs}{\end{eqnarray}}

\begin{document}
\tighten
\draft

\def\thefootnote{\fnsymbol{footnote}}

\baselineskip 6.0mm

\vspace{4mm}
                 
\begin{center}

{\Large \bf Exact Chromatic Polynomials for Toroidal Chains of Complete
Graphs}

\vspace{8mm}

\setcounter{footnote}{0}
Shu-Chiuan Chang\footnote{email: shu-chiuan.chang@sunysb.edu}

\vspace{6mm}

\ C. N. Yang Institute for Theoretical Physics  \\
State University of New York       \\
Stony Brook, N. Y. 11794-3840  \\

\vspace{10mm}

{\bf Abstract}
\end{center}

We present exact calculations of the partition function of the zero-temperature
Potts antiferromagnet (equivalently, the chromatic polynomial) for graphs of
arbitrarily great length composed of repeated complete subgraphs $K_b$ with
$b=5,6$ which have periodic or twisted periodic boundary condition in the
longitudinal direction. In the $L_x \to \infty$ limit, the continuous
accumulation set of the chromatic zeros ${\cal B}$ is determined. We give some
results for arbitrary $b$ including the extrema of the eigenvalues with
coefficients of degree $b-1$ and the explicit forms of some classes of
eigenvalues. We prove that the maximal point where ${\cal B}$ crosses the real
axis, $q_c$, satisfies the inequality $q_c \le b$ for $2 \le b$, the minimum
value of $q$ at which ${\cal B}$ crosses the real $q$ axis is $q=0$, and we
make a conjecture concerning the structure of the chromatic polynomial for 
Klein bottle strips.

\vspace{16mm}

\pagestyle{empty}
\newpage

\pagestyle{plain}
\pagenumbering{arabic}
\renewcommand{\thefootnote}{\arabic{footnote}}
\setcounter{footnote}{0}

\section{Introduction}

The $q$-state Potts antiferromagnet (AF) \cite{potts,wurev} exhibits nonzero
ground state entropy, $S_0 > 0$ (without frustration) for sufficiently large
$q$ on a given lattice $\Lambda$ or, more generally, on a graph $G$.  This is
equivalent to a ground state degeneracy per site $W > 1$, since $S_0 = k_B \ln
W$.  There is a close connection with graph theory here, since the
zero-temperature partition function of the above-mentioned $q$-state Potts
antiferromagnet on a graph $G$ satisfies

\beq 
Z(G,q,T=0)_{PAF}=P(G,q) \ ,
\label{zp} 
\eeq 
where $P(G,q)$ is the chromatic polynomial expressing the number of ways
of coloring the vertices of the graph $G$ with $q$ colors such that no two
adjacent vertices have the same color (for reviews, see 
\cite{rrev}-\cite{bbook}).  The minimum number of colors necessary for
such a coloring of $G$ is called the chromatic number, $\chi(G)$.  Thus

\beq 
W(\{G\},q) = \lim_{n \to \infty} P(G,q)^{1/n} \ ,
\label{w} 
\eeq 
where $n$ is the number of vertices of $G$ and $\{G\} = \lim_{n \to 
\infty}G$.  Where no confusion will result, we shall sometimes write $G$ rather
than $\{G\}$ for the infinite-length limit of a given type of strip graph. 
At certain special points $q_s$ (typically $q_s=0,1,..,\chi(G)$), one has 
the noncommutativity of limits 

\beq 
\lim_{q \to q_s} \lim_{n \to \infty} P(G,q)^{1/n} \ne \lim_{n \to \infty}
\lim_{q \to q_s}P(G,q)^{1/n} \ ,
\label{wnoncom} 
\eeq 
and hence it is necessary to specify the order of the limits in the
definition of $W(\{G\},q_s)$ \cite{w}. Denoting $W_{qn}$ and $W_{nq}$ as
the functions defined by the different order of limits on the left and
right-hand sides of (\ref{wnoncom}), we take $W \equiv W_{qn}$ here; this
has the advantage of removing certain isolated discontinuities that are
present in $W_{nq}$. 

Using the expression for $P(G,q)$, one can generalize $q$ from ${\mathbb
Z}_+$ to ${\mathbb C}$.  The zeros of $P(G,q)$ in the complex $q$ plane
are called chromatic zeros; a subset of these may form an accumulation set
in the $n \to \infty$ limit, denoted ${\cal B}$, which is the continuous
locus of points where $W(\{G\},q)$ is nonanalytic. 
\footnote{\footnotesize{For some families of graphs ${\cal B}$ may be
null, and $W$ may also be nonanalytic at certain discrete points.}} The
maximal region in the complex $q$ plane to which one can analytically
continue the function $W(\{G\},q)$ from physical values where there is
nonzero ground state entropy is denoted $R_1$.  The maximal value of $q$
where ${\cal B}$ intersects the (positive) real axis is labeled
$q_c(\{G\})$.  This point is important since $W(\{G\},q)$ is a real
analytic function from large values of $q$ down to $q_c(\{G\})$.

Here we present exact calculations of the partition function of the chromatic
polynomial $P(G,q)$ for graphs of arbitrarily great length composed of repeated
complete subgraphs $K_b$ with $b=5,6$ which have periodic or twisted periodic
boundary conditions in the longitudinal direction.\footnote{$K_n$ denotes the
complete graph, i.e. the graph with $n$ vertices such that each vertex is
connected by an edge to all of the other vertices.}  Thus, consider $L_x$
copies of the complete graph $K_{L_y}$ with vertex set $V(K_{L_y}) =
\{1,2,...,L_y\}$. Denote the edges joining each adjacent pair of $K_{L_y}$ and
$K^\prime_{L_y}$ graphs as $L=\{v_iv^\prime_j|i,j=1,...,L_y\}$, where
$v_i,v^\prime_j$ are vertices of adjacent complete graphs $K_{L_y}$ and
$K^\prime_{L_y}$, and impose periodic boundary conditions in the
longitudinal direction, $x$. In \cite{lse9908}, the graph with $L_x$
copies of $K_{L_y}$ for arbitrary $L_x$ and $L_y$ was called the bracelet
strip. The total number of vertices is $n = L_xL_y$ for these strips.

A generic form for chromatic polynomials for recursively defined families
of graphs (families of graphs that can be constructed via repeated 
addition of some subgraph), of which strip graphs $G_s$ are special cases, 
is \cite{bkw}

\beq 
P((G_s)_m,q) = \sum_{j=1}^{N_{G_s,\lambda}}
c_{G_s,j}(q)(\lambda_{G_s,j}(q))^{L_x} \ ,
\label{pgsum} 
\eeq 
where $c_{G_s,j}(q)$ and the $N_{G_s,\lambda}$ terms $\lambda_{G_s,j}(q)$
depend on the type of strip graph $G_s$ but are independent of $L_x$.

For a given type of strip graph $G_s$, we denote the sum of the
coefficients $c_{G_s,j}$ as

\beq
C(G_s,q)=\sum_{j=1}^{N_{G_s,\lambda}} c_{G_s,j}(q) \ .
\label{cgsum}
\eeq

Some works on calculations of chromatic polynomials for recursive families
of graphs include \cite{bds}-\cite{tor4}.

\section{Family of toroidal Strips}

Specifically, we consider the set of edges linking two successive complete
graphs on the chain is $L = \{11,22,33,...,L_yL_y\}$, which will be
denoted as a torus. For $L_y=2$, this is the ladder graph with vertex set
$T_2=K_2$ \cite{bds}. For the graph with $L_y=3, 4$, which has vertex set
$C_3=K_3$ and $K_4$, the chromatic polynomial was given in \cite{tk} and
\cite{dn}, respectively. In \cite{cprsg,amcp1}, these families of graphs
with linking edge sets $\{11,22,33,...,bb\}$ were denoted as $B_m(b)$,
where $b=L_y$, $m=L_x$. The corresponding Klein bottle strip is the same
as the toroidal strip except one of the linking edge set is 
$\{1L_y,2(L_y-1),...L_y1\}$. The form of the chromatic polynomials for the
$B_m(b)$ strips was determined to be 
\beq 
P(B_m(b),q) = \sum_{d=0}^b\sum_{j=1}^{N_{B_m(b),d,\lambda}}
c_{B_m(b),d,j}(q)c^\prime_{B_m(b),d,j}(\lambda_{B_m(b),d,j}(q))^m \ ,
\label{pbsum} 
\eeq 
where $d$ is the degree of the coefficient $c_{B_m(b),d,j}(q)$ as a
polynomial in $q$, and the $c^\prime_{B_m(b),d,j}$'s are numbers which
depend on $b$.  The structural feature that these coefficients can be
grouped into sets of fixed degree $d$ is similar to that found in
\cite{cf} for cyclic strips of the square and triangular lattice, although
for these bracelet chain graphs, for $b \ge 3$, there is more than one
coefficient of a given degree (see \cite{tk} for the case $b=3$), whereas
in contrast, for the cyclic strip graphs studied in \cite{cf} (and the
self-dual family studied in \cite{dg})  there is only one coefficient of
each degree.  The degree $d$ is denoted as the `level' in
\cite{lse0004,cprsg,amcp1}. The $\lambda_{B_m(b),d,j}(q)$ is the
eigenvalue of an appropriately defined transfer matrix \cite{matmeth} and
is a polynomial in $q$ of degree $b-d$. For sufficiently large integer 
$q$, the coefficient $c_{B_m(b),d,j}(q)c^\prime_{B_m(b),d,j}$ can be
interpreted as the multiplicity of this eigenvalue. 

There are partitions associated with each $d$ which were used to determine
the forms of $c_{B_m(b),d,j}$ in \cite{amcp1}. For a partition $X$ of $d$
such that $x_d + x_{d-1} + ... + x_1 = d$ and $x_d \le x_{d-1} \le ... \le 
x_1$ where $x_i$ for $2 \le i \le d$ are non-negative integers and $x_1$
is a positive integer, there is an associated partition $Y$ of $d(d+1)/2$
such
that $y_d + y_{d-1} + ... + y_1 = d(d+1)/2$ and $y_i = x_i + d - i$ for $1
\le i \le d$. Then the $c_{B_m(b),d,j}$ for a partition $X$ can be written
as 
\beq
c_{X,d} = \frac{1}{d!} \prod _{i=1}^d (q-y_i) \ .
\label{cx}
\eeq
For the partitions $[d]$ and $[1^d]$, the associated 
$\lambda_{B_m(b),d,j}(q)$'s were given in \cite{cprsg}. 

\subsection{$L_{\lowercase{y}}=5$ toroidal strip}

By using theorems in \cite{cprsg}, we find the chromatic polynomial for
$B_m(5)$, i.e., the $5 \times L_x$ strips with each transverse slice
forming a $K_5$ graph, and with edge linking set $L=\{11,22,33,44,55\}$ between
two adjacent slices. 

It is convenient to express the $\lambda$'s in term of the following
functions \cite{lse0004}
\beq
f_i(b,q) = \sum _{s=0}^{b-i} (-1)^s {b-i \choose s} (q-i-s)^{(b-i-s)} \ ,
\label{fi}
\eeq
where $q^{(k)}$ is falling factorial defined as
\beq
q^{(k)} = \prod _{s=0}^{k-1}(q-s) \ .
\label{falling}
\eeq
Here we adopt the convention that $f_i(b,q)=0$ if $b < i$.

For the case $b=5$, one has 
\beqs
f_0(5,q) & = & q^5 - 15q^4 + 95q^3 - 325q^2 + 609q - 501 \\
f_1(5,q) & = & q^4 - 14q^3 + 77q^2 - 200q + 209 \\
f_2(5,q) & = & q^3 - 12q^2 + 50q - 73 \\
f_3(5,q) & = & q^2 - 9q + 21 \\
f_4(5,q) & = & q - 5 \\
f_5(5,q) & = & 1 \ ,
\eeqs
and the chromatic polynomial for the $B_m(5)$ strip is 
\beq
P(B_m(5),q) = \sum_{d=0}^5\sum_{j=1}^{N_{B_m(5),d,\lambda}}
c_{B_m(5),d,j}(q)c^\prime_{B_m(5),d,j}(\lambda_{B_m(5),d,j}(q))^m \ ,
\label{pb5sum}
\eeq
where $N_{B_m(5),d,\lambda}$'s are equal to $1,2,5,9,9,1$ for $d$ from $0$
to $5$.  We find that the total number of distinct $\lambda$'s is
\beq
N_{P,B_m(5),\lambda} = \sum_{d=0}^5N_{B_m(5),d,\lambda} = 27 \ .
\label{nptotb5}
\eeq
We note that our result (\ref{nptotb5}) differs from the pattern
$N_{P,B_m(b),\lambda} = 2^b$ that was found for $1 \le b \le 4$, namely
$N_{P,B_m(1),\lambda}=2$, $N_{P,B_m(2),\lambda}=2^2$ \cite{bds},
$N_{P,B_m(3),\lambda}=2^3$ \cite{tk}, and $N_{P,B_m(4),\lambda}=2^4$
\cite{dn}. 

We calculate that the eigenvalue with coefficient of degree 0 in $q$ is 
unique and is given by 
\beq
\lambda_{B_m(5),0,1} = f_0(5,q) = q^5-15q^4+95q^3-325q^2+609q-501 \ .
\label{b5lam1}
\eeq

For the eigenvalues with coefficients of degree 1 in $q$ we find 
\beq 
\lambda_{B_m(5),1,1} = -f_1(5,q)+4f_2(5,q) = -q^4+18q^3-125q^2+400q-501
\label{b5lam2} 
\eeq
\beq 
\lambda_{B_m(5),1,2} = -f_1(5,q)-f_2(5,q) = -(q-4)(q^3-9q^2+29q-34) \ .
\label{b5lam3} 
\eeq

Next, for the eigenvalues with coefficients of degree 2 in $q$ we obtain 
\beq 
\lambda_{B_m(5),2,1} = f_2(5,q)-6f_3(5,q)+6f_4(5,q) = q^3-18q^2+110q-229
\label{b5lam4} 
\eeq
\beq
\lambda_{B_m(5),2,2} = f_2(5,q)-f_3(5,q)-4f_4(5,q) = q^3-13q^2+55q-74
\label{b5lam5}
\eeq
\beq
\lambda_{B_m(5),2,3} = f_2(5,q)+2f_3(5,q)+2f_4(5,q) = q^3-10q^2+34q-41
\label{b5lam6}
\eeq
\beq
\lambda_{B_m(5),2,4} = f_2(5,q)-3f_3(5,q) = q^3-15q^2+77q-136
\label{b5lam7}
\eeq
\beq
\lambda_{B_m(5),2,5} = f_2(5,q)+2f_3(5,q) = q^3-10q^2+32q-31 \ .
\label{b5lam8}
\eeq
For $1 \le j \le 3$ these $\lambda_{B_m(5),2,j}$'s correspond to the 
partition $[2]$ and for $j=4,5$ to the partition $[11]$.  

Proceeding to the eigenvalues with coefficients of degree 3, we find 
\beq
\lambda_{B_m(5),3,1} = -f_3(5,q)+6f_4(5,q)-6f_5(5,q) = -q^2+15q-57
\label{b5lam9}
\eeq
\beq
\lambda_{B_m(5),3,2} = -f_3(5,q)+f_4(5,q)+4f_5(5,q) = -q^2+10q-22
\label{b5lam10}
\eeq
\beq
\lambda_{B_m(5),3,3} = -f_3(5,q)-2f_4(5,q)-2f_5(5,q) = -q^2+7q-13
\label{b5lam11}
\eeq
\beq
\lambda_{B_m(5),3,4} = -f_3(5,q)+4f_4(5,q)-2f_5(5,q) = -q^2+13q-43
\label{b5lam12}
\eeq
\beq
\lambda_{B_m(5),3,5} = -f_3(5,q)+f_4(5,q)+f_5(5,q) = -(q-5)^2
\label{b5lam13}
\eeq
\beq
\lambda_{B_m(5),3,6} = -f_3(5,q)-f_4(5,q)+3f_5(5,q) = -q^2+8q-13
\label{b5lam14}
\eeq
\beq
\lambda_{B_m(5),3,7} = -f_3(5,q)-3f_4(5,q)-3f_5(5,q) = -(q-3)^2
\label{b5lam15}
\eeq
\beq
\lambda_{B_m(5),3,8} = -f_3(5,q)+2f_4(5,q) = -q^2+11q-31
\label{b5lam16}
\eeq
\beq
\lambda_{B_m(5),3,9} = -f_3(5,q)-3f_4(5,q) = -q^2+6q-6 \ .
\label{b5lam17}
\eeq
Here the $\lambda_{B_m(5),3,j}$ for $1 \le j \le 3$ correspond to the 
partition $[3]$, for $4 \le j \le 7$ to the partition $[21]$, and for 
$j=8,9$ to the partition $[111]$.  

For the eigenvalues with coefficients of degree 4 in $q$, we calculate 
\beq
\lambda_{B_m(5),4,1} = f_4(5,q)-4f_5(5,q) = q-9
\label{b5lam18}
\eeq
\beq
\lambda_{B_m(5),4,2} = f_4(5,q)+f_5(5,q) = q-4              
\label{b5lam19}
\eeq
\beq
\lambda_{B_m(5),4,3} = f_4(5,q)-3f_5(5,q) = q-8
\label{b5lam20}
\eeq
\beq
\lambda_{B_m(5),4,4} = f_4(5,q)+3f_5(5,q) = q-2              
\label{b5lam21}
\eeq
\beq
\lambda_{B_m(5),4,5} = f_4(5,q)+2f_5(5,q) = q-3              
\label{b5lam22}
\eeq
\beq
\lambda_{B_m(5),4,6} = f_4(5,q) = q-5              
\label{b5lam23}
\eeq
\beq
\lambda_{B_m(5),4,7} = f_4(5,q)-2f_5(5,q) = q-7             
\label{b5lam24}
\eeq
\beq
\lambda_{B_m(5),4,8} = f_4(5,q)-f_5(5,q) = q-6              
\label{b5lam25}
\eeq
\beq
\lambda_{B_m(5),4,9} = f_4(5,q)+4f_5(5,q) = q-1 \ .
\label{b5lam26}
\eeq
Here the $\lambda_{B_m(5),4,j}$ for $j=1,2$ correspond to the partition $[4]$,
for $j=3,5,6$ to the partition $[31]$, for $j=4,6,7$ to the partition $[211]$,
for $j=5,7$ to the partition $[22]$, and for $j=8,9$ to the partition
$[1111]$. Notice that the terms $\lambda_{B_m(5),4,j}$ for $5 \le j \le 7$
involve degeneracies from two different associated partitions. 

Finally, for the eigenvalue with coefficient of degree 5, we have 
\beq
\lambda_{B_m(5),5,1} = -f_5(5,q) = -1 \ .
\label{b5lam27}
\eeq
This involves a degeneracy from all the possible partitions.  

The coefficient $c_{B_m(5),d,j}(q)c^\prime_{B_m(5),d,j}$ of degree 0 in
$q$ is
\beq
c_{B_m(5),0,1}c^\prime_{B_m(5),0,1}= 1 \ .
\label{b5c1}
\eeq

The coefficients of degree 1 in $q$ are
\beq
c_{B_m(5),1,1}c^\prime_{B_m(5),1,1}= q-1
\label{b5c2}   
\eeq
\beq
c_{B_m(5),1,2}c^\prime_{B_m(5),1,2}= 4(q-1) \ ,
\label{b5c3}
\eeq
where $c_{B_m(5),1,j}=q-1$ for $j=1,2$, and $c^\prime_{B_m(5),1,j}=1,4$
for $j=1,2$. 

For $d=2$, there are partitions $[2]$ and $[11]$, and the associated
$c_{X,2}$ are $c_{[2],2}=\frac{1}{2}q(q-3)$ and 
$c_{[11],2}=\frac{1}{2}(q-1)(q-2)$. The coefficients of degree 2 in $q$
are listed in Table \ref{b5d2}

\begin{table}
\caption{\footnotesize{Coefficients of degree 2 in $q$ for $B_m(5)$ 
strip.}}
\begin{center}
\begin{tabular}{|c|c|c|c|}
$j$ & partition & $c^\prime_{B_m(5),2,j}$ & 
$c_{B_m(5),2,j}c^\prime_{B_m(5),2,j}$ \\ \hline\hline
1  & [2] & 1 & $\frac{1}{2}q(q-3)$ \\ \hline
2  & [2] & 4 & $2q(q-3)$ \\ \hline
3  & [2] & 5 & $\frac{5}{2}q(q-3)$ \\ \hline
4  & [11] & 4 & $2(q-1)(q-2)$ \\ \hline
5  & [11] & 6 & $3(q-1)(q-2)$ \\ 
\end{tabular}
\end{center}
\label{b5d2}
\end{table} 

For $d=3$, there are partitions $[3]$, $[21]$ and $[111]$, and the 
associated $c_{X,3}$ are $c_{[3],3}=\frac{1}{6}q(q-1)(q-5)$,
$c_{[21],3}=\frac{1}{6}q(q-2)(q-4)$ and  
$c_{[111],3}=\frac{1}{6}(q-1)(q-2)(q-3)$. The coefficients of degree 3 in
$q$ are listed in Table \ref{b5d3}

\begin{table}
\caption{\footnotesize{Coefficients of degree 3 in $q$ for $B_m(5)$ 
strip.}}
\begin{center}
\begin{tabular}{|c|c|c|c|}
$j$ & partition & $c^\prime_{B_m(5),3,j}$ & 
$c_{B_m(5),3,j}c^\prime_{B_m(5),3,j}$ \\ \hline\hline
1  & [3] & 1 & $\frac{1}{6}q(q-1)(q-5)$ \\ \hline
2  & [3] & 4 & $\frac{2}{3}q(q-1)(q-5)$ \\ \hline
3  & [3] & 5 & $\frac{5}{6}q(q-1)(q-5)$ \\ \hline
4  & [21] & 8 & $\frac{4}{3}q(q-2)(q-4)$ \\ \hline
5  & [21] & 10 & $\frac{5}{3}q(q-2)(q-4)$ \\ \hline
6  & [21] & 12 & $2q(q-2)(q-4)$ \\ \hline
7  & [21] & 10 & $\frac{5}{3}q(q-2)(q-4)$ \\ \hline
8  & [111] & 6 & $(q-1)(q-2)(q-3)$ \\ \hline
9  & [111] & 4 & $\frac{2}{3}(q-1)(q-2)(q-3)$ \\ 
\end{tabular}
\end{center}
\label{b5d3}
\end{table} 

For $d=4$, there are partitions $[4]$, $[31]$, $[22]$, $[211]$ and
$[1111]$, and the associated $c_{X,4}$ are 
$c_{[4],4}=\frac{1}{24}q(q-1)(q-2)(q-7)$, 
$c_{[31],4}=\frac{1}{24}q(q-1)(q-3)(q-6)$,
$c_{[22],4}=\frac{1}{24}q(q-1)(q-4)(q-5)$,
$c_{[211],4}=\frac{1}{24}q(q-2)(q-3)(q-5)$, and
$c_{[1111],4}=\frac{1}{24}(q-1)(q-2)(q-3)(q-4)$. The coefficients of
degree 4 in $q$ are listed in Table \ref{b5d4}

\begin{table}
\caption{\footnotesize{Coefficients of degree 4 in $q$ for $B_m(5)$ 
strip.}}
\begin{center}
\begin{tabular}{|c|c|c|c|}
$j$ & partition & $c^\prime_{B_m(5),4,j}$ & 
$c_{B_m(5),4,j}c^\prime_{B_m(5),4,j}$ \\ \hline\hline
1  & [4] & 1 & $\frac{1}{24}q(q-1)(q-2)(q-7)$ \\ \hline
2  & [4] & 4 & $\frac{1}{6}q(q-1)(q-2)(q-7)$ \\ \hline
3  & [31] & 12 & $\frac{1}{2}q(q-1)(q-3)(q-6)$ \\ \hline
4  & [211] & 12 & $\frac{1}{2}q(q-2)(q-3)(q-5)$ \\ \hline
5  & [31] & 18 & $\frac{1}{6}q(q-1)(7q^2-63q+131)$ \\ 
   & [22] & 10 &  \\ \hline
6  & [31] & 15 & $\frac{5}{4}q(q-3)(q^2-7q+8)$ \\ 
   & [211] & 15 &  \\ \hline
7  & [211] & 18 & $\frac{1}{6}q(q-5)(7q^2-35q+37)$ \\ 
   & [22] & 10 &  \\ \hline
8  & [1111] & 4 & $\frac{1}{6}(q-1)(q-2)(q-3)(q-4)$ \\ \hline
9  & [1111] & 1 & $\frac{1}{24}(q-1)(q-2)(q-3)(q-4)$ \\
\end{tabular}
\end{center}
\label{b5d4}
\end{table} 

Finally, the coefficient of degree 5 in $q$ is given by
\beq 
c_{B_m(5),5,1}c^\prime_{B_m(5),5,1}= q^5-15q^4+75q^3-145q^2+89q-1 \ .
\label{b5c27}
\eeq

The sum of all the coefficients is equal to 
\beqs
C(B_m(5),q) & = & \sum_{d=0}^5\sum_{j=1}^{N_{B_m(5),d,\lambda}}
c_{B_m(5),d,j}(q)c^\prime_{B_m(5),d,j} \nonumber \\
 & = & q(q-1)(q-2)(q-3)(q-4) = P(K_5,q) \ .
\label{cb5sum}
\eeqs 
The chromatic number is $\chi (B_m(5)) = 5$.

\begin{figure}[hbtp]
\centering
\leavevmode
\epsfxsize=4.0in
\begin{center}
\leavevmode
\epsffile{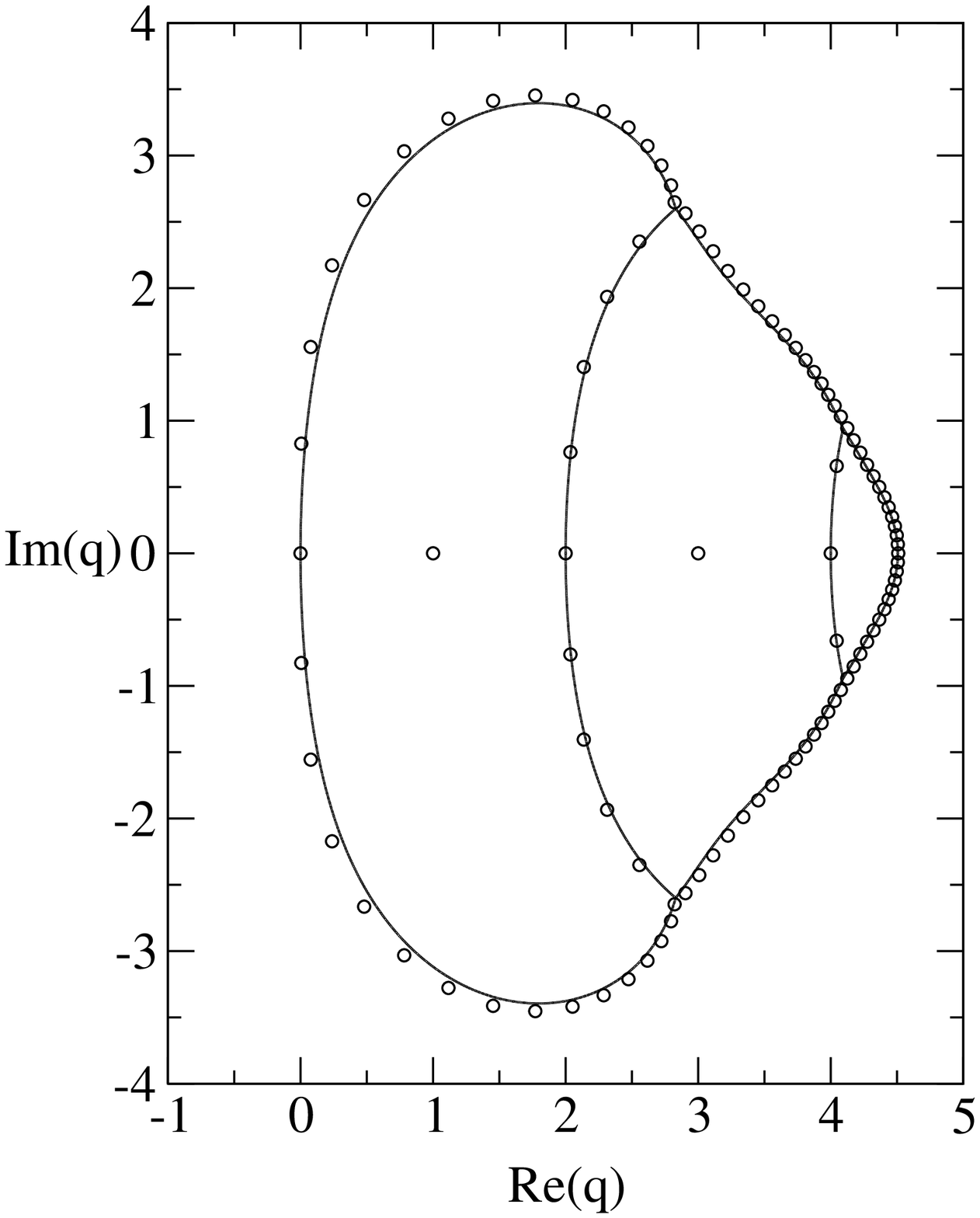}
\end{center}
\caption{\footnotesize{Locus ${\cal B}$ for the $m \to \infty$ limit of the
family $B_m(5)$ with toroidal boundary conditions and chromatic zeros for 
$B_m(5)$ with $m=20$ (i.e., $n=100$).}}
\label{sqkpxpy5}
\end{figure}

The locus ${\cal B}$ and chromatic zeros for the $B_m(5)$ strip with
$L_x=m=20$ are shown in Fig. \ref{sqkpxpy5}.  The locus ${\cal B}$ crosses
the real $q$-axis at $q=0,2,4$ and $q_c(B_m(5))$, where

\beq
q_c(B_m(5)) \simeq 4.50634 \cdots \ .
\label{qcb5}
\eeq

The locus ${\cal B}$ has support for $Re(q) \ge 0$, and separates the $q$
plane into four regions.  The outermost one, region $R_1$, extends to
infinite $|q|$ and includes the intervals $q \ge q_c(B_m(5))$ and $q \le
0$ on the real $q$ axis.  Region $R_2$ includes the real interval $4 \le q
\le q_c(B_m(5))$, region $R_3$ includes the real interval $2 \le q \le 4$,
while region $R_4$ includes the real interval $0 \le q \le 2$.  In regions
$R_i$, $1 \le i \le 4$, the dominant terms are $\lambda_{B_m(5),0,1}$,
$\lambda_{B_m(5),3,1}$, $\lambda_{B_m(5),2,1}$, and
$\lambda_{B_m(5),1,1}$, respectively. Thus, the $q_c(B_m(5))$
given in (\ref{qcb5}) is the degeneracy between $|\lambda_{B_m(5),0,1}|$
and $|\lambda_{B_m(5),3,1}|$, and is the real solution of
$q^5-15q^4+95q^3-326q^2+624q-558=0$.

\subsection{$L_{\lowercase{y}}=6$ toroidal strip}

We have also succeeded in obtaining the chromatic polynomial for $B_m(6)$,
i.e., the $6 \times L_x$ strips with each transverse slice forming a $K_6$
graph, and with edge linking set $L=\{11,22,33,44,55,66\}$ between two
adjacent
slices. 

For $b=6$, one has
\beqs
f_0(6,q) & = & q^6 - 21q^5 + 190q^4 - 965q^3 + 2944q^2 - 5155q + 4051 \\
f_1(6,q) & = & q^5 - 20q^4 + 165q^3 - 710q^2 + 1609q - 1546 \\
f_2(6,q) & = & q^4 - 18q^3 + 125q^2 - 400q + 501 \\
f_3(6,q) & = & q^3 - 15q^2 + 77q - 136 \\
f_4(6,q) & = & q^2 - 11q + 31 \\
f_5(6,q) & = & q - 6 \\
f_6(6,q) & = & 1 \ ,
\eeqs
and the chromatic polynomial for the $B_m(6)$ strip is
\beq
P((B_m(6),q) = \sum_{d=0}^6\sum_{j=1}^{N_{B_m(6),d,\lambda}}
c_{B_m(6),d,j}(q)c^\prime_{B_m(6),d,j}(\lambda_{B_m(6),d,j}(q))^m \ ,
\label{pb6sum}
\eeq
where $N_{B_m(6),d,\lambda}$'s are equal to $1,2,5,10,16,11,1$ for $d$
from $0$ to $6$.  We find that the total number of distinct $\lambda$'s is
\beq
N_{P,B_m(6),\lambda } = 46 \ .
\label{nptotb6}
\eeq

The eigenvalue with coefficient of degree 0 in $q$ is unique and is given
by
\beq
\lambda_{B_m(6),0,1} = f_0(6,q) =
q^6-21q^5+190q^4-965q^3+2944q^2-5155q+4051 \ .
\label{b6lam1}
\eeq

For the eigenvalues with coefficients of degree 1 in $q$ we have
\beq 
\lambda_{B_m(6),1,1} = -f_1(6,q)+5f_2(6,q) =
-q^5+25q^4-255q^3+1335q^2-3609q+4051
\label{b6lam2} 
\eeq
\beq 
\lambda_{B_m(6),1,2} = -f_1(6,q)-f_2(6,q) =
-(q-5)(q^4-14q^3+77q^2-200q+209) \ .
\label{b6lam3} 
\eeq

For the eigenvalues with coefficients of degree 2 in $q$ we find
\beq 
\lambda_{B_m(6),2,1} = f_2(6,q)-8f_3(6,q)+12f_4(6,q) =
q^4-26q^3+257q^2-1148q+1961
\label{b6lam4} 
\eeq
\beq
\lambda_{B_m(6),2,2} = f_2(6,q)-2f_3(6,q)-6f_4(6,q) =
q^4-20q^3+149q^2-488q+587
\label{b6lam5}
\eeq
\beq
\lambda_{B_m(6),2,3} = f_2(6,q)+2f_3(6,q)+2f_4(6,q) =
q^4-16q^3+97q^2-268q+291
\label{b6lam6}
\eeq
\beq
\lambda_{B_m(6),2,4} = f_2(6,q)-4f_3(6,q) = q^4-22q^3+185q^2-708q+1045
\label{b6lam7}
\eeq
\beq
\lambda_{B_m(6),2,5} = f_2(6,q)+2f_3(6,q) = q^4-16q^3+95q^2-246q+229 \ .
\label{b6lam8}
\eeq
For $1 \le j \le 3$, these $\lambda_{B_m(6),2,j}$ correspond to the 
partition $[2]$ and for $j=4,5$ to the partition $[11]$.

For the eigenvalues with coefficients of degree 3 in $q$ we obtain
\beq
\lambda_{B_m(6),3,1} = -f_3(6,q)+9f_4(6,q)-18f_5(6,q)+6f_6(6,q) =
-q^3+24q^2-194q+529
\label{b6lam9}
\eeq
\beq
\lambda_{B_m(6),3,2} = -f_3(6,q)+3f_4(6,q)+6f_5(6,q)-6f_6(6,q) =
-q^3+18q^2-104q+187
\label{b6lam10}
\eeq
\beq
\lambda_{B_m(6),3,3} = -f_3(6,q)-f_4(6,q)+2f_5(6,q)+6f_6(6,q) =
-q^3+14q^2-64q+99
\label{b6lam11}
\eeq
\beq
\lambda_{B_m(6),3,4} = -f_3(6,q)-3f_4(6,q)-6f_5(6,q)-6f_6(6,q) =
-q^3+12q^2-50q+73
\label{b6lam12}
\eeq
\beq
\lambda_{B_m(6),3,5} = -f_3(6,q)+6f_4(6,q)-6f_5(6,q) = -q^3+21q^2-149q+358
\label{b6lam13}
\eeq
\beq
\lambda_{B_m(6),3,6} = -f_3(6,q)+2f_4(6,q)+2f_5(6,q) = -(q-6)(q^2-11q+31)
\label{b6lam14}
\eeq
\beq
\lambda_{B_m(6),3,7} = -f_3(6,q)+6f_5(6,q) = -q^3+15q^2-71q+100
\label{b6lam15}
\eeq
\beq
\lambda_{B_m(6),3,8} = -f_3(6,q)-3f_4(6,q)-3f_5(6,q) = -q^3+12q^2-47q+61
\label{b6lam16}
\eeq
\beq
\lambda_{B_m(6),3,9} = -f_3(6,q)+3f_4(6,q) = -q^3+18q^2-110q+229
\label{b6lam17}
\eeq
\beq
\lambda_{B_m(6),3,10} = -f_3(6,q)-3f_4(6,q) = -q^3+12q^2-44q+43 \ .
\label{b6lam18}
\eeq
Here the $\lambda_{B_m(6),3,j}$ for $1 \le j \le 4$ correspond to the
partition $[3]$, for $5 \le j \le 8$ to the partition $[21]$, and for
$j=9,10$ to the partition $[111]$.

Proceeding to the eigenvalues with coefficients of degree 4 in $q$ we
calculate
\beq
\lambda_{B_m(6),4,1} = f_4(6,q)-8f_5(6,q)+12f_6(6,q) = q^2-19q+91
\label{b6lam19}
\eeq
\beq
\lambda_{B_m(6),4,2} = f_4(6,q)-2f_5(6,q)-6f_6(6,q) = q^2-13q+37              
\label{b6lam20}
\eeq
\beq
\lambda_{B_m(6),4,3} = f_4(6,q)+2f_5(6,q)+2f_6(6,q) = q^2-9q+21
\label{b6lam21}
\eeq
\beq
\lambda_{B_m(6),4,4} = f_4(6,q)-6f_6(6,q) = q^2-11q+25              
\label{b6lam22}
\eeq
\beq
\lambda_{B_m(6),4,5} = f_4(6,q)-2f_5(6,q)-2f_6(6,q) = q^2-13q+41              
\label{b6lam23}
\eeq
\beq
\lambda_{B_m(6),4,6} = f_4(6,q) = q^2-11q+31              
\label{b6lam24}
\eeq
\beq
\lambda_{B_m(6),4,7} = f_4(6,q)+3f_5(6,q)+3f_6(6,q) = (q-4)^2             
\label{b6lam25}
\eeq
\beq
\lambda_{B_m(6),4,8} = f_4(6,q)-6f_5(6,q)+6f_6(6,q) = q^2-17q+73              
\label{b6lam26}
\eeq
\beq
\lambda_{B_m(6),4,9} = f_4(6,q)+f_5(6,q)-3f_6(6,q) = q^2-10q+22
\label{b6lam27}
\eeq
\beq
\lambda_{B_m(6),4,10} = f_4(6,q)+4f_5(6,q)+6f_6(6,q) = q^2-7q+13
\label{b6lam28}
\eeq
\beq
\lambda_{B_m(6),4,11} = f_4(6,q)-4f_5(6,q)+2f_6(6,q) = q^2-15q+57
\label{b6lam29}
\eeq
\beq
\lambda_{B_m(6),4,12} = f_4(6,q)+2f_5(6,q)-4f_6(6,q) = q^2-9q+15
\label{b6lam30}
\eeq
\beq
\lambda_{B_m(6),4,13} = f_4(6,q)-f_5(6,q)-f_6(6,q) = (q-6)^2
\label{b6lam31}
\eeq
\beq
\lambda_{B_m(6),4,14} = f_4(6,q)+4f_5(6,q)+4f_6(6,q) = q^2-7q+11
\label{b6lam32}
\eeq
\beq
\lambda_{B_m(6),4,15} = f_4(6,q)-2f_5(6,q) = q^2-13q+43
\label{b6lam33}
\eeq
\beq
\lambda_{B_m(6),4,16} = f_4(6,q)+4f_5(6,q) = q^2-7q+7 \ .
\label{b6lam34}
\eeq
Here the $\lambda_{B_m(6),4,j}$ for $1 \le j \le 3$ correspond to the
partition $[4]$, for $4 \le j \le 8$ to the partition $[31]$, for $9 \le j
\le 11$ to the partition $[22]$, for $11 \le j \le 14$ to the partition
$[211]$, and for $j=15,16$ to the partition $[1111]$. Notice that the term
$\lambda_{B_m(6),4,11}$ involves a degeneracy from two different
associated
partitions.

For the eigenvalues with coefficients of degree 5 in $q$ we find
\beq
\lambda_{B_m(6),5,1} = -f_5(6,q)+5f_6(6,q) = -q+11
\label{b6lam35}
\eeq
\beq
\lambda_{B_m(6),5,2} = -f_5(6,q)-f_6(6,q) = -q+5
\label{b6lam36}
\eeq
\beq
\lambda_{B_m(6),5,3} = -f_5(6,q)+4f_6(6,q) = -q+10
\label{b6lam37}
\eeq
\beq
\lambda_{B_m(6),5,4} = -f_5(6,q) = -q+6
\label{b6lam38}
\eeq
\beq
\lambda_{B_m(6),5,5} = -f_5(6,q)-2f_6(6,q) = -q+4
\label{b6lam39}
\eeq
\beq
\lambda_{B_m(6),5,6} = -f_5(6,q)+3f_6(6,q) = -q+9
\label{b6lam40}
\eeq
\beq
\lambda_{B_m(6),5,7} = -f_5(6,q)+f_6(6,q) = -q+7
\label{b6lam41}
\eeq
\beq
\lambda_{B_m(6),5,8} = -f_5(6,q)-3f_6(6,q) = -q+3
\label{b6lam42}
\eeq
\beq
\lambda_{B_m(6),5,9} = -f_5(6,q)+2f_6(6,q) = -q+8
\label{b6lam43}
\eeq
\beq
\lambda_{B_m(6),5,10} = -f_5(6,q)-4f_6(6,q) = -q+2
\label{b6lam44}
\eeq
\beq
\lambda_{B_m(6),5,11} = -f_5(6,q)-5f_6(6,q) = -q+1 \ .
\label{b6lam45}
\eeq
Here $\lambda_{B_m(6),5,j}$ for $j=1,2$ correspond to the partition $[5]$,
for $3 \le j \le 5$ to the partition $[41]$, for $5 \le j \le 7$ to the
partition $[32]$, for $j=4,6,8$ to the partition $[311]$, for $j=2,8,9$ to
the partition $[221]$, for $j=4,9,10$ to the partition $[2111]$, and for
$j=7,11$ to the partition $[11111]$. Notice that the terms
$\lambda_{B_m(6),5,j}$ for $j=2$ and $5 \le j \le 9$ involve degeneracies
from two different associated partitions, and for $j=4$ from three
associated partitions. 

Finally, for the eigenvalue with coefficient of degree 6 in $q$ we have 
\beq
\lambda_{B_m(6),6,1} = f_6(6,q) = 1 \ .
\label{b6lam46}
\eeq
This involves a degeneracy from all the possible partitions.  

The coefficient $c_{B_m(6),d,j}(q)c^\prime_{B_m(6),d,j}$ of degree 0 in
$q$ is
\beq
c_{B_m(6),0,1}c^\prime_{B_m(6),0,1}= 1 \ .
\label{b6c1}
\eeq

The coefficients of degree 1 in $q$ are
\beq
c_{B_m(6),1,1}c^\prime_{B_m(6),1,1}= q-1
\label{b6c2}   
\eeq
\beq
c_{B_m(6),1,2}c^\prime_{B_m(6),1,2}= 5(q-1) \ ,
\label{b6c3}
\eeq
where $c_{B_m(6),1,j}=q-1$ for $j=1,2$, and $c^\prime_{B_m(6),1,j}=1,5$
for $j=1,2$.

The coefficients of degree 2 in $q$ are listed in Table \ref{b6d2}

\begin{table}
\caption{\footnotesize{Coefficients of degree 2 in $q$ for $B_m(6)$ 
strip.}}
\begin{center}
\begin{tabular}{|c|c|c|c|}
$j$ & partition & $c^\prime_{B_m(6),2,j}$ & 
$c_{B_m(6),2,j}c^\prime_{B_m(6),2,j}$ \\ \hline\hline
1  & [2] & 1 & $\frac{1}{2}q(q-3)$ \\ \hline
2  & [2] & 5 & $\frac{5}{2}q(q-3)$ \\ \hline
3  & [2] & 9 & $\frac{9}{2}q(q-3)$ \\ \hline
4  & [11] & 5 & $\frac{5}{2}(q-1)(q-2)$ \\ \hline
5  & [11] & 10 & $5(q-1)(q-2)$ \\ 
\end{tabular}
\end{center}
\label{b6d2}
\end{table} 

The coefficients of degree 3 in $q$ are listed in Table \ref{b6d3}

\begin{table}
\caption{\footnotesize{Coefficients of degree 3 in $q$ for $B_m(6)$ 
strip.}}
\begin{center}
\begin{tabular}{|c|c|c|c|}
$j$ & partition & $c^\prime_{B_m(6),3,j}$ & 
$c_{B_m(6),3,j}c^\prime_{B_m(6),3,j}$ \\ \hline\hline
1  & [3] & 1 & $\frac{1}{6}q(q-1)(q-5)$ \\ \hline
2  & [3] & 5 & $\frac{5}{6}q(q-1)(q-5)$ \\ \hline
3  & [3] & 9 & $\frac{3}{2}q(q-1)(q-5)$ \\ \hline
4  & [3] & 5 & $\frac{5}{6}q(q-1)(q-5)$ \\ \hline
5  & [21] & 10 & $\frac{5}{3}q(q-2)(q-4)$ \\ \hline
6  & [21] & 18 & $3q(q-2)(q-4)$ \\ \hline
7  & [21] & 20 & $\frac{10}{3}q(q-2)(q-4)$ \\ \hline
8  & [21] & 32 & $\frac{16}{3}q(q-2)(q-4)$ \\ \hline
9  & [111] & 10 & $\frac{5}{3}(q-1)(q-2)(q-3)$ \\ \hline
10  & [111] & 10 & $\frac{5}{3}(q-1)(q-2)(q-3)$ \\ 
\end{tabular}
\end{center}
\label{b6d3}
\end{table} 

The coefficients of degree 4 in $q$ are listed in Table \ref{b6d4}

\begin{table}
\caption{\footnotesize{Coefficients of degree 4 in $q$ for $B_m(6)$ 
strip.}}
\begin{center}
\begin{tabular}{|c|c|c|c|}
$j$ & partition & $c^\prime_{B_m(6),4,j}$ & 
$c_{B_m(6),4,j}c^\prime_{B_m(6),4,j}$ \\ \hline\hline
1  & [4] & 1 & $\frac{1}{24}q(q-1)(q-2)(q-7)$ \\ \hline
2  & [4] & 5 & $\frac{5}{24}q(q-1)(q-2)(q-7)$ \\ \hline
3  & [4] & 9 & $\frac{3}{8}q(q-1)(q-2)(q-7)$ \\ \hline
4  & [31] & 30 & $\frac{5}{4}q(q-1)(q-3)(q-6)$ \\ \hline
5  & [31] & 27 & $\frac{9}{8}q(q-1)(q-3)(q-6)$ \\ \hline
6  & [31] & 15 & $\frac{5}{8}q(q-1)(q-3)(q-6)$ \\ \hline
7  & [31] & 48 & $2q(q-1)(q-3)(q-6)$ \\ \hline
8  & [31] & 15 & $\frac{5}{8}q(q-1)(q-3)(q-6)$ \\ \hline
9  & [22] & 32 & $\frac{4}{3}q(q-1)(q-4)(q-5)$ \\ \hline
10  & [22] & 10 & $\frac{5}{12}q(q-1)(q-4)(q-5)$ \\ \hline
11  & [22] & 18 & $\frac{1}{2}q(q-5)(2q-3)(2q-7)$ \\ 
    & [211] & 30 &  \\ \hline
12  & [211] & 30 & $\frac{5}{4}q(q-2)(q-3)(q-5)$ \\ \hline
13  & [211] & 48 & $2q(q-2)(q-3)(q-5)$ \\ \hline
14  & [211] & 27 & $\frac{9}{8}q(q-2)(q-3)(q-5)$ \\ \hline
15  & [1111] & 10 & $\frac{5}{12}(q-1)(q-2)(q-3)(q-4)$ \\ \hline
16  & [1111] & 5 & $\frac{5}{24}(q-1)(q-2)(q-3)(q-4)$ \\
\end{tabular}
\end{center}
\label{b6d4}
\end{table} 

For $d=5$, there are partitions $[5]$, $[41]$, $[32]$, $[311]$, $[221]$,
$[2111]$ and $[11111]$, and the associated $c_{X,5}$ are 
$c_{[5],5}=\frac{1}{120}q(q-1)(q-2)(q-3)(q-9)$, 
$c_{[41],5}=\frac{1}{120}q(q-1)(q-2)(q-4)(q-8)$,
$c_{[32],5}=\frac{1}{120}q(q-1)(q-2)(q-5)(q-7)$,
$c_{[311],5}=\frac{1}{120}q(q-1)(q-3)(q-4)(q-7)$,
$c_{[221],5}=\frac{1}{120}q(q-1)(q-3)(q-5)(q-6)$,
$c_{[2111],5}=\frac{1}{120}q(q-2)(q-3)(q-4)(q-6)$, and
$c_{[11111],5}=\frac{1}{120}(q-1)(q-2)(q-3)(q-4)(q-5)$. We list the
coefficients of degree 5 in $q$ in Table \ref{b6d5}

\begin{table}
\caption{\footnotesize{Coefficients of degree 5 in $q$ for $B_m(6)$ 
strip.}}
\begin{center}
\begin{tabular}{|c|c|c|c|}
$j$ & partition & $c^\prime_{B_m(6),5,j}$ & 
$c_{B_m(6),5,j}c^\prime_{B_m(6),5,j}$ \\ \hline\hline
1  & [5] & 1 & $\frac{1}{120}q(q-1)(q-2)(q-3)(q-9)$ \\ \hline
2  & [5] & 5 & $\frac{1}{4}q(q-1)(q-3)(q-4)(q-7)$ \\ 
   & [221] & 25 &  \\ \hline
3  & [41] & 20 & $\frac{1}{6}q(q-1)(q-2)(q-4)(q-8)$ \\ \hline
4  & [41] & 36 & $\frac{1}{5}q(q-4)(7q^3-77q^2+217q-162)$ \\
   & [311] & 96 &  \\ 
   & [2111] & 36 &  \\ \hline
5  & [41] & 40 & $q(q-1)(q-2)(q^2-12q+34)$ \\ 
   & [32] & 80 &  \\ \hline
6  & [32] & 45 & $\frac{1}{8}q(q-1)(q-7)(7q^2-49q+78)$ \\
   & [311] & 60 &  \\ \hline
7  & [32] & 25 & $\frac{1}{4}(q-1)(q-2)(q-5)(q^2-7q+2)$ \\
   & [11111] & 5 &  \\ \hline
8  & [311] & 60 & $\frac{1}{8}q(q-1)(q-3)(7q^2-77q+202)$ \\
   & [221] & 45 &  \\ \hline
9  & [221] & 80 & $q(q-3)(q-6)(q^2-6q+6)$ \\ 
   & [2111] & 40 &  \\ \hline
10  & [2111] & 20 & $\frac{1}{6}q(q-2)(q-3)(q-4)(q-6)$ \\ \hline
11  & [11111] & 1 & $\frac{1}{120}(q-1)(q-2)(q-3)(q-4)(q-5)$ \\ 
\end{tabular}
\end{center}
\label{b6d5}
\end{table} 

Finally, the coefficient of degree 6 is
\beq c_{B_m(6),6,1}c^\prime_{B_m(6),6,1}=
q^6-21q^5+160q^4-545q^3+814q^2-415q+1 \ .
\label{b6c46}
\eeq

The sum of all the coefficients is equal to 
\beqs
C(B_m(6),q) & = & \sum_{d=0}^6\sum_{j=1}^{N_{B_m(6),d,\lambda}}
c_{B_m(6),d,j}(q)c^\prime_{B_m(6),d,j} \nonumber \\
 & = & q(q-1)(q-2)(q-3)(q-4)(q-5) = P(K_6,q) \ .
\label{cb6sum}
\eeqs
The chromatic number is $\chi (B_m(6)) = 6$.

\begin{figure}[hbtp]
\centering
\leavevmode
\epsfxsize=4.0in
\begin{center}
\leavevmode
\epsffile{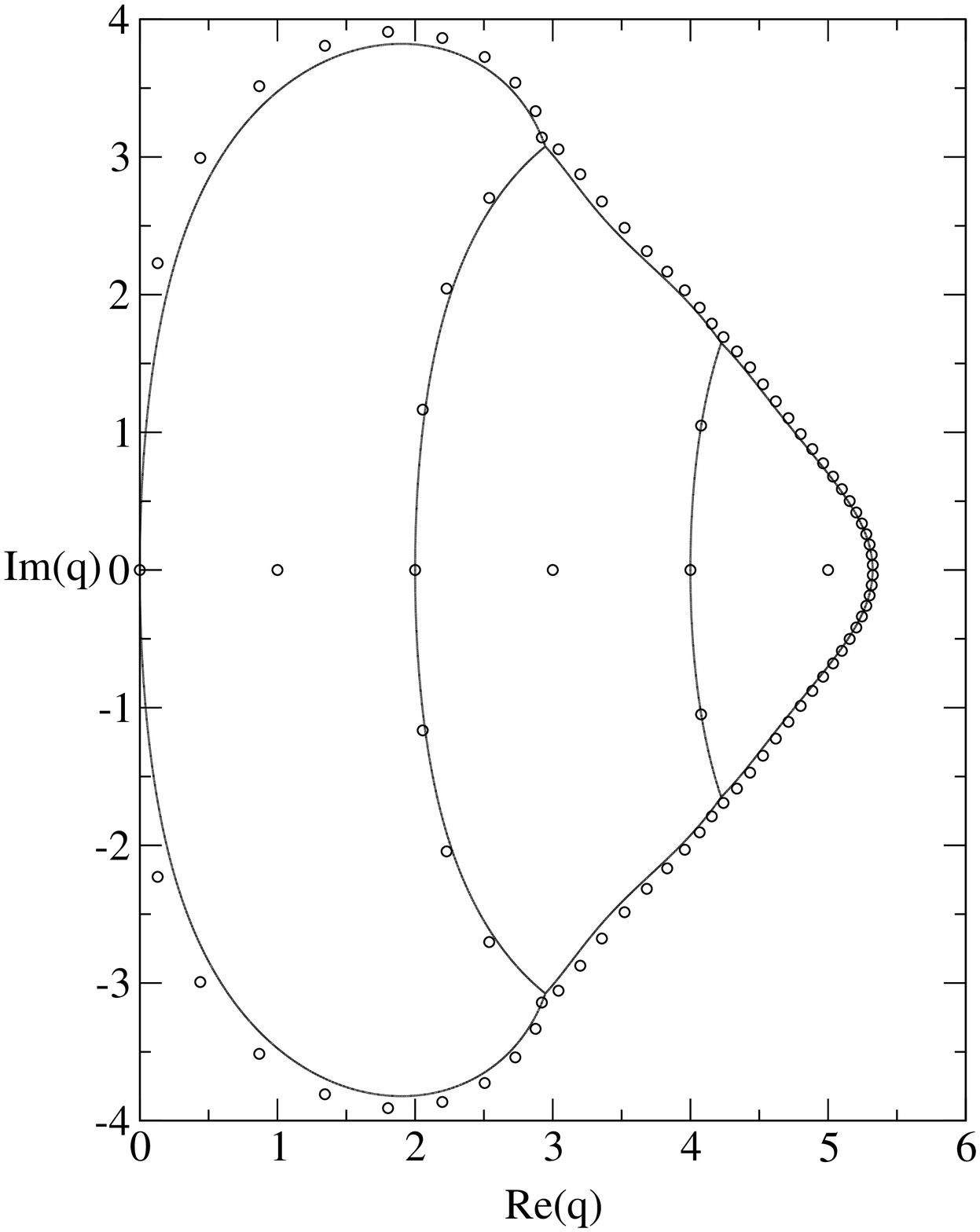}
\end{center}
\caption{\footnotesize{Locus ${\cal B}$ for the $m \to \infty$ limit of the
family $B_m(6)$ with toroidal boundary conditions and chromatic zeros for 
$B_m(6)$ with $m=15$ (i.e., $n=90$).}}
\label{sqkpxpy6}
\end{figure}

The locus ${\cal B}$ and chromatic zeros for the $B_m(6)$ strip with
$L_x=m=15$ are shown in Fig. \ref{sqkpxpy6}.  The locus ${\cal B}$ crosses
the real $q$-axis at $q=0,2,4$ and $q_c(B_m(6))$, where

\beq
q_c(B_m(6)) \simeq 5.3236 \cdots \ .
\label{qcb6}
\eeq

The locus ${\cal B}$ has support for $Re(q) \ge 0$, and separates the $q$
plane into four regions.  The outermost one, region $R_1$, extends to
infinite $|q|$ and includes the intervals $q \ge q_c(B_m(6))$ and $q \le
0$ on the real $q$ axis.  Region $R_2$ includes the real interval $4 \le q
\le q_c(B_m(6))$, region $R_3$ includes the real interval $2 \le q \le 4$,
while region $R_4$ includes the real interval $0 \le q \le 2$. In regions
$R_i$, $1 \le i \le 4$, the dominant terms are $\lambda_{B_m(6),0,1}$, 
$\lambda_{B_m(6),3,1}$, $\lambda_{B_m(6),2,1}$, and 
$\lambda_{B_m(6),1,1}$, respectively. Thus, the $q_c(B_m(6))$
given in (\ref{qcb6}) is the degeneracy between $|\lambda_{B_m(6),0,1}|$
and $|\lambda_{B_m(6),3,1}|$, and is the real solution of
$q^5-19q^4+152q^3-660q^2+1600q-1761=0$.

\section{Family of Klein bottle Strips}

Consider the graph with $L_x$ set of $K_{b}$ with $L_x-1$ edge linking 
sets $\{11,22,...,bb\}$ and one edge linking set 
$\{1b,2(b-1),...,b1\}$. This is the
Klein bottle strip and will be denoted as ${\bar B}_m(b)$. The chromatic
polynomials has the same form as given in eq. (\ref{pbsum}), and the
$\lambda$'s are the same as those of the corresponding toroidal strip 
with the same $b$. The only difference is the coefficients.

\subsection{$L_{\lowercase{y}}=5$ Klein bottle strip}

For $b=5$ considered here, the coefficients $c_{{\bar
B}_m(5),d,j}(q)c^\prime_{{\bar B}_m(5),d,j}$ are summarized in Table
\ref{kb5}

\begin{table}
\caption{\footnotesize{Coefficients for $\bar B_m(5)$ strip.}}
\begin{center}
\begin{tabular}{|c|c|c|c|c|}
$d$ & $j$ & partition & $c^\prime_{\bar B_m(5),d,j}$ & 
$c_{\bar B_m(5),d,j}c^\prime_{\bar B_m(5),d,j}$ \\ \hline\hline
0 & 1  & [0] & 1 & 1 \\ \hline\hline
1 & 1  & [1] & 1 & $q-1$ \\ \hline
1 & 2  & [1] & 0 & 0 \\ \hline\hline
2 & 1  & [2] & 1 & $\frac{1}{2}q(q-3)$ \\ \hline
2 & 2  & [2] & 0 & 0 \\ \hline
2 & 3  & [2] & 1 & $\frac{1}{2}q(q-3)$ \\ \hline
2 & 4  & [11] & 0 & 0 \\ \hline
2 & 5  & [11] & $-2$ & $-(q-1)(q-2)$ \\ \hline\hline
3 & 1  & [3] & 1 & $\frac{1}{6}q(q-1)(q-5)$ \\ \hline
3 & 2  & [3] & 0 & 0 \\ \hline
3 & 3  & [3] & 1 & $\frac{1}{6}q(q-1)(q-5)$ \\ \hline
3 & 4  & [21] & 0 & 0 \\ \hline
3 & 5  & [21] & 2 & $\frac{1}{3}q(q-2)(q-4)$ \\ \hline
3 & 6  & [21] & $-4$ & $-\frac{2}{3}q(q-2)(q-4)$ \\ \hline
3 & 7  & [21] & 2 & $\frac{1}{3}q(q-2)(q-4)$ \\ \hline
3 & 8  & [111] & $-2$ & $-\frac{1}{3}(q-1)(q-2)(q-3)$ \\ \hline
3 & 9  & [111] & 0 & 0 \\ \hline\hline
4 & 1  & [4] & 1 & $\frac{1}{24}q(q-1)(q-2)(q-7)$ \\ \hline
4 & 2  & [4] & 0 & 0 \\ \hline
4 & 3  & [31] & 0 & 0 \\ \hline
4 & 4  & [211] & 0 & 0 \\ \hline
4 & 5  & [31] & $-6$ & $-\frac{1}{6}q(q-1)(q^2-9q+17)$ \\ 
  &    & [22] & 2 &  \\ \hline
4 & 6  & [31] & 3 & $\frac{1}{4}q(q-3)(q^2-7q+8)$ \\ 
  &    & [211] & 3 &  \\ \hline
4 & 7  & [211] & $-6$ & $-\frac{1}{6}q(q-5)(q^2-5q+7)$ \\ 
  &    & [22] & 2 &  \\ \hline
4 & 8  & [1111] & 0 & 0 \\ \hline
4 & 9  & [1111] & 1 & $\frac{1}{24}(q-1)(q-2)(q-3)(q-4)$ \\ \hline\hline
5 & 1  & see text & & $q-1$ \\
\end{tabular}
\end{center}
\label{kb5}
\end{table} 
\noindent where for $d=5$, it involves all possible partitions with
appropriate $c^\prime_{\bar B_m(5),5,j}$ which are not listed here. Notice
that the coefficients for $\lambda_{{\bar B}_m(5),d,j}$ with
$(d,j)=(1,2)$, $(2,2)$, $(2,4)$, $(3,2)$, $(3,4)$, $(3,9)$, $(4,2)$,
$(4,3)$, $(4,4)$, $(4,8)$ are zero, i.e., these ten $\lambda$'s do not
contribute to the chromatic polynomial of the Klein bottle strip. We thus
find that the total number of distinct $\lambda$'s for the $b=5$ Klein
bottle strip with indicated linking is 
\beq
N_{P,{\bar B}_m(5),\lambda } = 17 \ . 
\label{nptotb5klein}
\eeq

The sum of all of the coefficients for the Klein bottle strip is 
\beq
C({\bar B}_m(5),q)=\sum_{d=0}^5\sum_{j=1}^{N_{{\bar B}_m(5),d,\lambda}} 
c_{{\bar B}_m(5),d,j}(q)c^\prime_{{\bar B}_m(5),d,j} =
0 \ ,
\label{cb5ksum}
\eeq
which is easy to understand by the coloring argument.

\subsection{$L_{\lowercase{y}}=6$ Klein bottle strip}

For $b=6$, the coefficients $c_{{\bar B}_m(6),d,j}(q)c^\prime_{{\bar
B}_m(6),d,j}$ are summarized in Table \ref{kb6}

\begin{table}
\caption{\footnotesize{Coefficients for $\bar B_m(6)$ strip.}}
\begin{center}
\begin{tabular}{|c|c|c|c|c|}
$d$ & $j$ & partition & $c^\prime_{\bar B_m(6),d,j}$ &
$c_{\bar B_m(6),d,j}c^\prime_{\bar B_m(6),d,j}$ \\ \hline\hline
0 & 1  & [0] & 1 & 1 \\ \hline\hline
1 & 1  & [1] & 1 & $q-1$ \\ \hline
1 & 2  & [1] & $-1$ & $-(q-1)$ \\ \hline\hline 
2 & 1  & [2] & 1 & $\frac{1}{2}q(q-3)$ \\ \hline
2 & 2  & [2] & $-1$ & $-\frac{1}{2}q(q-3)$ \\ \hline
2 & 3  & [2] & 3 & $\frac{3}{2}q(q-3)$ \\ \hline
2 & 4  & [11] & $-1$ & $-\frac{1}{2}(q-1)(q-2)$ \\ \hline
2 & 5  & [11] & $-2$ & $-(q-1)(q-2)$ \\ \hline\hline
3 & 1  & [3] & 1 & $\frac{1}{6}q(q-1)(q-5)$ \\ \hline
3 & 2  & [3] & $-1$ & $-\frac{1}{6}q(q-1)(q-5)$ \\ \hline
3 & 3  & [3] & 3 & $\frac{1}{2}q(q-1)(q-5)$ \\ \hline
3 & 4  & [3] & $-3$ & $-\frac{1}{2}q(q-1)(q-5)$ \\ \hline
3 & 5  & [21] & $-2$ & $-\frac{1}{3}q(q-2)(q-4)$ \\ \hline
3 & 6  & [21] & 6 & $q(q-2)(q-4)$ \\ \hline
3 & 7  & [21] & $-4$ & $-\frac{2}{3}q(q-2)(q-4)$ \\ \hline
3 & 8  & [21] & 0 & 0 \\ \hline
3 & 9  & [111] & $-2$ & $-\frac{1}{3}(q-1)(q-2)(q-3)$ \\ \hline
3 & 10  & [111] & 2 & $\frac{1}{3}(q-1)(q-2)(q-3)$ \\ \hline\hline
4 & 1  & [4] & 1 & $\frac{1}{24}q(q-1)(q-2)(q-7)$ \\ \hline
4 & 2  & [4] & $-1$ & $-\frac{1}{24}q(q-1)(q-2)(q-7)$ \\ \hline
4 & 3  & [4] & 3 & $\frac{1}{8}q(q-1)(q-2)(q-7)$ \\ \hline
4 & 4  & [31] & $-6$ & $-\frac{1}{4}q(q-1)(q-3)(q-6)$ \\ \hline
4 & 5  & [31] & 9 & $\frac{3}{8}q(q-1)(q-3)(q-6)$ \\ \hline
4 & 6  & [31] & $-9$ & $-\frac{3}{8}q(q-1)(q-3)(q-6)$ \\ \hline
4 & 7  & [31] & 0 & 0 \\ \hline
4 & 8  & [31] & $-3$ & $-\frac{1}{8}q(q-1)(q-3)(q-6)$ \\ \hline
4 & 9  & [22] & 0 & 0 \\ \hline
4 & 10  & [22] & 6 & $\frac{1}{4}q(q-1)(q-4)(q-5)$ \\ \hline
4 & 11  & [22] & 6 & $-\frac{1}{2}q(q-5)$ \\ 
  &     & [211] & $-6$ &  \\ \hline
4 & 12  & [211] & 6 & $\frac{1}{4}q(q-2)(q-3)(q-5)$ \\ \hline
4 & 13  & [211] & 0 & 0 \\ \hline
4 & 14  & [211] & $-9$ & $-\frac{3}{8}q(q-2)(q-3)(q-5)$ \\ \hline
4 & 15  & [1111] & 2 & $\frac{1}{12}(q-1)(q-2)(q-3)(q-4)$ \\ \hline
4 & 16  & [1111] & 1 & $\frac{1}{24}(q-1)(q-2)(q-3)(q-4)$ \\ \hline\hline
5 & 1  & [5] & 1 & $\frac{1}{120}q(q-1)(q-2)(q-3)(q-9)$ \\ \hline
5 & 2  & [5] & $-1$ & $\frac{1}{60}q(q-1)(q-3)(7q^2-77q+216)$ \\ 
  &    & [221] & 15 &  \\ \hline
5 & 3  & [41] & $-4$ & $-\frac{1}{30}q(q-1)(q-2)(q-4)(q-8)$ \\ \hline
5 & 4  & [41] & 12 & $-q(q-2)(q-4)$ \\
  &    & [2111] & $-12$ &  \\ \hline
5 & 5  & [41] & $-8$ & $-\frac{1}{15}q(q-1)(q-2)(q-4)(q-8)$ \\ \hline
5 & 6  & [32] & 15 & $\frac{1}{40}q(q-1)(q-7)(q^2-7q+2)$ \\
  &    & [311] & $-12$ &  \\ \hline
5 & 7  & [32] & $-15$ & $-\frac{1}{60}(q-1)(q-2)(q-5)(7q^2-49q-6)$ \\
  &    & [11111] & 1 &  \\ \hline
5 & 8  & [311] & 12 & $-\frac{1}{40}q(q-1)(q-3)(q^2-11q+38)$ \\
  &    & [221] & $-15$ &  \\ \hline
5 & 9  & [2111] & 8 & $\frac{1}{15}q(q-2)(q-3)(q-4)(q-6)$ \\ \hline
5 & 10  & [2111] & 4 & $\frac{1}{30}q(q-2)(q-3)(q-4)(q-6)$ \\ \hline
5 & 11  & [11111] & $-1$ & $-\frac{1}{120}(q-1)(q-2)(q-3)(q-4)(q-5)$ \\
\hline\hline
6 & 1  & see text & & $-1$ \\
\end{tabular}
\end{center}
\label{kb6}
\end{table}

\noindent where for $d=6$, it involves all possible partitions with
appropriate $c^\prime_{\bar B_m(6),6,j}$ which are not listed here. Notice
that the coefficients for $\lambda_{{\bar B}_m(6),d,j}$ with
$(d,j)=(3,8)$, $(4,7)$, $(4,9)$, $(4,13)$ are zero, i.e., these four
$\lambda$'s do not contribute to the chromatic polynomial of the Klein
bottle strip. We thus find that the total number of distinct $\lambda$'s
for the $b=6$ Klein bottle strip with indicated linking is 
\beq
N_{P,{\bar B}_m(6)} = 42 \ . 
\label{nptotb6klein}
\eeq

The sum of all of the coefficients for the Klein bottle
strip is again zero,
\beq
C({\bar B}_m(6),q)=\sum_{d=0}^6\sum_{j=1}^{N_{{\bar B}_m(6),d,\lambda}} 
c_{{\bar B}_m(6),d,j}(q)c^\prime_{{\bar B}_m(6),d,j} = 0 \ .
\label{cb6ksum}
\eeq

\section{Properties of eigenvalues, coefficients, and ${\cal B}$}

We first list the number of eigenvalues, $N_{B_m(b),d,\lambda}$, for $1
\le b \le 6$, $0 \le d \le b$, the total number of eigenvalues,
$N_{P,B_m(b),\lambda}$, and the points at which ${\cal B}$ crosses the
real $q$ axis, as well as the corresponding values for the Klein bottle
strips in Table \ref{proptable}.

\begin{table}
\caption{\footnotesize{Properties of $P$ and ${\cal B}$ for strip graphs
$B_m(b)$ and $\bar B_m(b)$. The properties apply for a given strip of size
$L_y \times L_x$; some apply for arbitrary $L_x$, such as
$N_{P,B_m(b),\lambda}$, while others apply for the infinite-length limit,
such as the properties of the locus ${\cal B}$. For the boundary
conditions in the $y$ and $x$ directions ($BC_y$, $BC_x$), P, and T denote
periodic, and orientation-reversed 
(twisted) periodic. $N_{B_m(b),d,\lambda}$'s are abbreviated as 
$N_{d,\lambda}$, $N_{P,B_m(b),\lambda}$ is abbreviated as $N_{P,\lambda}$,
and similarly for $\bar B_m(b)$ strips. The blank entries are zero. The
column denoted BCR lists the points at which ${\cal B}$ crosses the real
$q$ axis; the largest of these is $q_c(B_m(b))$ for the given family
$B_m(b)$ or $\bar B_m(b)$. Column labelled ``SN'' refers to whether ${\cal
B}$ has \underline{s}upport for \underline{n}egative $Re(q)$, indicated as
yes (y) or no (n).}}
\begin{center}
\begin{tabular}{|c|c|c|c|c|c|c|c|c|c|c|c|c|}
$b$ & $BC_y$ & $BC_x$ & $N_{0,\lambda}$ & $N_{1,\lambda}$ &
$N_{2,\lambda}$ & $N_{3,\lambda}$ & $N_{4,\lambda}$ & $N_{5,\lambda}$ &
$N_{6,\lambda}$ & $N_{P,\lambda}$ & BCR & SN \\ \hline\hline
1  & P & P & 1 & 1 &  &  &  &  &  & 2 & 0, 2 & n \\ \hline
2  & P & P & 1 & 2 & 1 &  &  &  &  & 4 & 0, 2 & n \\ \hline
2  & P & TP & 1 & 2 & 1 &  &  &  &  & 4 & 0, 2 & n \\ \hline
3  & P & P & 1 & 2 & 4 & 1 &  &  &  & 8 & 0, 2, 3 & n \\ \hline
3  & P & TP & 1 & 1 & 2 & 1 &  &  &  & 5 & 0, 2, 3 & n \\ \hline
4  & P & P & 1 & 2 & 5 & 7 & 1 &  &  & 16 & 0, 2, 3.67 & n \\ \hline
4  & P & TP & 1 & 2 & 5 & 7 & 1 &  &  & 16 & 0, 2, 3.67 & n \\ \hline
5  & P & P & 1 & 2 & 5 & 9 & 9 & 1 &  & 27 & 0, 2, 4, 4.51 & n \\ \hline
5  & P & TP & 1 & 1 & 3 & 6 & 5 & 1 &  & 17 & 0, 2, 4, 4.51 & n \\ \hline
6  & P & P & 1 & 2 & 5 & 10 & 16 & 11 & 1 & 46 & 0, 2, 4, 5.32 & n \\
\hline
6  & P & TP & 1 & 2 & 5 & 9 & 13 & 11 & 1 & 42 & 0, 2, 4, 5.32 & n \\ 
\end{tabular}
\end{center}
\label{proptable}
\end{table} 

We conjecture the following $q$ value at which ${\cal B}$ crosses the
real axis for the $B_m(b)$ strips,
\beqs
\quad {\cal B} \supset \{q=0, \ 2, \ 4, ..., 2[\frac{b-1}{2}], q_c(B_m(b)) 
\} \ .
\label{bcrossq}
\eeqs
where $[\nu]$ denotes the integral part of $\nu$, and the dominant
eigenvalue in interval between $2(d^\prime -1)$ and $2d^\prime$ is
$\lambda_{B_m(b),d^\prime,1}$ for $1 \le d^\prime \le [\frac{b+1}{2} ]$.

\bigskip

The upper and lower bounds for the eigenvalues with coefficients of
degree $b-1$ can be determined in the following theorem.

\bigskip

{\bf Theorem 1} \quad The eigenvalues with coefficients of degree $b-1$
are bounded between $(-1)^{b-1} (q-1)$ and $(-1)^{b-1} (q-2b+1)$.

{\sl Proof}. \quad Denote the $b! \times b!$ matrix for the eigenvalues
with coefficients of degree $b-1$ as $M_{b-1}$. By theorem 2 of
\cite{cprsg}, the diagonal elements of $M_{b-1}$ are all the same to be 
$(-1)^{b-1} f_{b-1}(b,q) = (-1)^{b-1} (q-b)$, and other non-zero elements
of $M_{b-1}$ are all the same to be $(-1)^b f_b(b,q) = (-1)^b$. Now
consider $M^\prime _{b-1} = M_{b-1} - (-1)^{b-1} (q-b)I$, where $I$ is the
$b! \times b!$ identity matrix. The number of non-zero elements, $(-1)^b$, 
in every row and columns of $M^\prime _{b-1}$ is $b-1$ by the construction
of the matrix. If we denote the eigenvalues of $M^\prime _{b-1}$ as
$\lambda ^\prime$ and the normalized eigenvector of $M^\prime _{b-1}$ as
$W^T = (w_1,w_2,...,w_{b!})$, then we have
\beqs
& & M^\prime _{b-1} W = \lambda ^\prime W \cr\cr 
& \Rightarrow & |M^\prime _{b-1} W|^2 = |\lambda ^\prime W|^2 \cr\cr
& \Rightarrow & (b-1)(w_1^2+w_2^2+...w_{b!}^2) + 2\sum_{i,j,i\ne j}w_i w_j
= \lambda ^{\prime 2} (w_1^2+w_2^2+...w_{b!}^2) \cr\cr 
& \Rightarrow & \lambda ^{\prime 2} = b-1 + 2\sum_{i,j,i\ne j}w_i w_j \ ,
\label{theorem1a}
\eeqs
where we sum all the possible multiplications of two different $w_i$ and
$w_j$ in each row of $M^\prime _{b-1}$. Therefore,
\beqs
\lambda ^{\prime 2} & \le & b-1 +\sum_{i,j,i\ne j} (w_i^2+w_j^2) \cr\cr
& = & b-1 + (b-1)(b-2) = (b-1)^2 \ ,
\label{theorem1b}
\eeqs
that is, $-(b-1) \le \lambda ^\prime \le b-1$. The upper and lower bounds
for the eigenvalues of $M_{b-1}$ are
\beq
(-1)^{b-1}(q-b) \pm (b-1) = \cases{ (-1)^{b-1} (q-1) & \cr
                                   (-1)^{b-1} (q-2b+1) & } \ .
\label{theorem1c}
\eeq
It is easy to see that $(-1)^{b-1} (q-2b+1)$ is one of the principal terms 
and $(-1)^{b-1} (q-1)$ is one of the alternating terms given in 
\cite{cprsg}. It follows that these values are realized as the largest and
smallest values for the eigenvalues of $M_{b-1}$
\ $\Box$.

\bigskip

We shall show that the locus ${\cal B}$ crosses the real $q$ axis at
$q=0$, and doesn't cross any $q \in {\mathbb R}_{\ -}$. We need the
following lemma.

{\bf Lemma 1} \quad On the non-positive real $q$ axis, the dominant
eigenvalue with coefficient of degree $d$ is
\beq
\lambda _{B_m(b),d} = (-1)^d\sum _{j=0}^d (-1)^j {d \choose j} (b-d)^{(j)}
f_{d+j}(b,q) \ .
\label{lemma1a}
\eeq

{\sl Proof}. \quad The matrix for the eigenvalues with coefficients of
degree $d$ can be written as \cite{cprsg} $M_d = (-1)^d f_d(b,q)I +
(-1)^{d+1} f_{d+1}(b,q) M_{d,1} + (-1)^{d+2} f_{d+2}(b,q) M_{d,2} +
...$, where the non-zero elements are 1 in every $M_{d,j}$, and the number
of these in each row and column is ${d \choose j} (b-d)^{(j)}$. By the
same reason given in Theorem 1, the magnitudes of the eigenvalues of
$M_{d,j}$ are bounded above by ${d \choose j} (b-d)^{(j)}$. Since the
signs of $f_{d+j}(b,q)$ alternate for even $j$ and odd $j$ for any
non-positive $q$ value, $\lambda _{B_m(b),d}$ given in eq. (\ref{lemma1a})
is the upper bound for the eigenvalues with coefficients of degree
$d$. Since $\lambda _{B_m(b),d}$ is the eigenvalue with  
$c^\prime_{B_m(b),d,j} = 1$ given in \cite{cprsg}, the proof is completed.
\ $\Box$.

\bigskip

{\bf Theorem 2} \quad The minimum real $q$ at which ${\cal B}$ crosses
the real $q$ axis is $q=0$.

{\sl Proof}. \quad We will begin with the proof that $|\lambda
_{B_m(b),0}| > |\lambda _{B_m(b),1}| > ...$ on the whole negative real $q$
axis. The first few examples of $\lambda _{B_m(b),d}$'s are
\beqs
\lambda _{B_m(b),0} & = & f_0(b,q) \cr\cr
\lambda _{B_m(b),1} & = & -f_1(b,q) + (b-1)f_2(b,q) \cr\cr
\lambda _{B_m(b),2} & = & f_2(b,q) - 2(b-2)f_3(b,q) + 
(b-2)(b-3)f_4(b,q) \cr\cr
\lambda _{B_m(b),3} & = & -f_3(b,q) + 3(b-3)f_4(b,q) -
3(b-3)(b-4)f_5(b,q) + (b-3)(b-4)(b-5)f_6(b,q) \cr\cr
\lambda _{B_m(b),4} & = & f_4(b,q) - 4(b-4)f_5(b,q) +
6(b-4)(b-5)f_6(b,q) - 4(b-4)(b-5)(b-6)f_7(b,q) \cr\cr
& & + (b-4)(b-5)(b-6)(b-7)f_8(b,q) \ .
\label{theorem2a}
\eeqs
To show that $f_0(b,q)$ has the largest magnitude on the negative real $q$
axis, we only need $|f_i(b,q)| > |f_{i+1}(b,q) - (b-i-1)f_{i+2}(b,q)|$ as
follows,
\beqs
& & |f_i(b,q)| - |f_{i+1}(b,q) - (b-i-1)f_{i+2}(b,q)| \cr\cr
& = & (-1)^{b-i}f_i(b,q) - (-1)^{b-i-1} \biggl [ f_{i+1}(b,q) -
(b-i-1)f_{i+2}(b,q) \biggr ] \cr\cr
& = & (-1)^{b-i} \biggl [ \sum _{s=0}^{b-i} (-1)^s {b-i \choose s}
(q-i-s)^{(b-i-s)} + f_{i+1}(b,q) - (b-i-1)f_{i+2}(b,q) \biggr ] \cr\cr
& = & (-1)^{b-i} \biggl [ (q-i)^{(b-i)} + \sum _{s=0}^{b-i-1} (-1)^{s+1} 
{b-i \choose s+1} (q-i-s-1)^{(b-i-s-1)} \cr\cr
& & + \sum _{s=0}^{b-i-1} (-1)^s {b-i-1 \choose s} (q-i-1-s)^{(b-i-1-s)} -
(b-i-1)f_{i+2}(b,q) \biggr ] \cr\cr
& = & (-1)^{b-i} \biggl [ (q-i)^{(b-i)} + (-1)^{b-i} + (-1)^{b-i-1} + \sum
_{s=0}^{b-i-2} (-1)^{s+1} (q-i-1-s)^{(b-i-1-s)} \cr\cr
& & \times \biggl ( \frac{(b-i)!}{(s+1)!(b-i-s-1)!} - 
\frac{(b-i-1)!}{s!(b-i-1-s)!} \biggr ) - (b-i-1)f_{i+2}(b,q) \biggr ]
\cr\cr
& = & (-1)^{b-i} \biggl [ (q-i)^{(b-i)} + \sum _{s=0}^{b-i-2} (-1)^{s+1}
(q-i-1-s)^{(b-i-1-s)} {b-i-1 \choose s+1} \cr\cr
& & - (b-i-1) \sum _{s=0}^{b-i-2} (-1)^s {b-i-2 \choose s}
(q-i-2-s)^{(b-i-2-s)} \biggr ] \cr\cr
& = & (-1)^{b-i} \biggl [ (q-i)^{(b-i)} + \sum _{s=0}^{b-i-2} (-1)^{s+1} 
(q-i-2-s)^{(b-i-2-s)} \cr\cr 
& & \times \biggl [ (q-i-1-s) \frac{(b-i-1)!}{(s+1)!(b-i-2-s)!} + 
(b-i-1) \frac{(b-i-2)!}{s!(b-i-2-s)!} \biggr ] \biggr ] \cr\cr
& = & (-1)^{b-i} \biggl [ (q-i)^{(b-i)} + \sum _{s=0}^{b-i-2} (-1)^{s+1}
(q-i-2-s)^{(b-i-2-s)} (b-i-1) {b-i-2 \choose s} \frac{q-i}{s+1} \biggr ]
\cr\cr
& > & 0 \quad {\rm for} \ q < 0 \ .
\label{theorem2b}
\eeqs
For $q=0$, all above argument are still valid, with the exception that
$|f_i(b,0)| = |f_{i+1}(b,0) - (b-i-1)f_{i+2}(b,0)|$ when $i=0$,
i.e. $|\lambda _{B_m(b),0}| = |\lambda _{B_m(b),1}|$ for $q=0$.
\ $\Box$.

\bigskip

We will prove that the maximal value of $q$ where ${\cal B}$ intersects
the (positive) real axis is smaller than $q=b$ for $b \ge 2$. We have the
following lemma.

{\bf Lemma 2} \quad $f_i(b,q) \ge 0$ if $q \ge b$.

{\sl Proof}. \quad Since we have $(q-i-s)^{(b-i-s)} > 0$ when $q \ge b$,
one only need to show that the magnitudes of the terms in
eq. (\ref{fi}) decrease for $s$ from 0 to $b-i$.
\beqs
& & {b-i \choose s} (q-i-s)^{(b-i-s)} - {b-i \choose s+1}
(q-i-s-1)^{(b-i-s-1)} \cr\cr
& = & (q-i-s-1)^{(b-i-s-1)} \biggl [ \frac{(b-i)!}{s!(b-i-s)!} (q-i-s) -
\frac{(b-i)!}{(s+1)!(b-i-s-1)!} \biggr ] \cr\cr
& = & (q-i-s-1)^{(b-i-s-1)} \frac{(b-i)!}{(s+1)!(b-i-s)!} \biggl [
(s+1)(q-i-s) - (b-i-s) \biggr ] \cr\cr
& = & (q-i-s-1)^{(b-i-s-1)} \frac{(b-i)!}{(s+1)!(b-i-s)!} \biggl [
s(q-i-s) + q-b \biggr ] \cr\cr
& \ge & 0 \quad {\rm for} \ 0 \le s \le b-i-1 \ {\rm and} \ q \ge b \ ,
\label{lemma2a}
\eeqs
where the equal sign is realized only when $s=0$ and $q=b$. Therefore,
$f_i(b,q) = 0$ only when $i=b-1$ at $q=b$, and $f_i(b,q) > 0$ for any
other $0 \le i \le b$ and $q \ge b$.
\ $\Box$.

\bigskip

It is known that $q_c(B_m(1))=2$, $q_c(B_m(2))=2$ \cite{bds,w}, and
$q_c(B_m(3))=3$ \cite{tk}.  Thus, for $b=1$, $q_c=b+1$ while for $b=2$ and for
$b=3$, $q_c$ has the respective values $q_c=b$.  It is of interest to ask how
$q_c(B_m(b))$ is related to $b$ for $b \ge 4$.  We determine this relation with
the following theorem.

{\bf Theorem 3} \quad $q_c(B_m(b)) < b$ for $b \ge 4$.
 
{\sl Proof}. \quad By a reason similar to that given in Lemma 1 and the result
of Lemma 2, the magnitude of the eigenvalues with coefficients of degree $d$
are bounded above by
\beq
\lambda ^\prime _{B_m(b),d} = \sum _{j=0}^d {d \choose j} (b-d)^{(j)} 
f_{d+j}(b,q) \quad {\rm for} \ q \ge b \ ,
\label{theorem3a}
\eeq
where it is equal to $\lambda _{B_m(b),b-1,1}$ for $d=b-1$ at $q=b$ and
equal to $\lambda _{B_m(b),0,1}$ for $d=0$ and any $q$. The first few 
examples of $\lambda ^\prime _{B_m(b),d}$'s are
\beqs
\lambda ^\prime _{B_m(b),0} & = & f_0(b,q) \cr\cr
\lambda ^\prime _{B_m(b),1} & = & f_1(b,q) + (b-1)f_2(b,q) \cr\cr
\lambda ^\prime _{B_m(b),2} & = & f_2(b,q) + 2(b-2)f_3(b,q) +
(b-2)(b-3)f_4(b,q) \cr\cr
\lambda ^\prime _{B_m(b),3} & = & f_3(b,q) + 3(b-3)f_4(b,q) +
3(b-3)(b-4)f_5(b,q) + (b-3)(b-4)(b-5)f_6(b,q) \cr\cr
\lambda ^\prime _{B_m(b),4} & = & f_4(b,q) + 4(b-4)f_5(b,q) +
6(b-4)(b-5)f_6(b,q) + 4(b-4)(b-5)(b-6)f_7(b,q) \cr\cr
& & + (b-4)(b-5)(b-6)(b-7)f_8(b,q) \ .
\label{theorem3b}
\eeqs
To show that $f_0(b,q) = \lambda _{B_m(b),0,1}$ has the largest magnitude
for $q \ge b$, we first need $f_i(b,q) > f_{i+1}(b,q) + 
(b-i-1)f_{i+2}(b,q)$ for $0 \le i \le b-2$ and $q \ge b$,
\beqs
& & f_i(b,q) - f_{i+1}(b,q) - (b-i-1)f_{i+2}(b,q) \cr\cr
& = & \sum _{s=0}^{b-i} (-1)^s {b-i \choose s} (q-i-s)^{(b-i-s)} - \sum
_{s=0}^{b-i-1} (-1)^s {b-i-1 \choose s} (q-i-1-s)^{(b-i-1-s)} \cr\cr 
& & - (b-i-1)f_{i+2}(b,q) \cr\cr
& = & (-1)^{b-i} + \sum _{s=0}^{b-i-1} (-1)^s (q-i-1-s)^{(b-i-1-s)}
\frac{(b-i-1)!}{s!(b-i-s)!} \biggl [ (b-i)(q-i-s) - (b-i-s) \biggr ] 
\cr\cr 
& & - (b-i-1)f_{i+2}(b,q) \cr\cr
& = & (-1)^{b-i} + \sum _{s=0}^{b-i-1} (-1)^s (q-i-1-s)^{(b-i-1-s)} \biggl
[ {b-i-1 \choose s} (b-i-1) + {b-i \choose s} (q-b) \biggr ] \cr\cr 
& & - (b-i-1)f_{i+2}(b,q) \cr\cr
& = & (-1)^{b-i} + (-1)^{b-i-1} [b-i-1+(b-i)(q-b)] + \sum _{s=0}^{b-i-2}
(-1)^s (q-i-2-s)^{(b-i-2-s)} \cr\cr
& & \times \biggl [ (q-i-1-s) \biggl [ {b-i-1 \choose s} (b-i-1) + {b-i
\choose s} (q-b) \biggr ] - (b-i-1) {b-i-2 \choose s} \biggr ] \cr\cr
& = & (-1)^{b-i} + (-1)^{b-i-1} [b-i-1+(b-i)(q-b)] + \sum _{s=0}^{b-i-2}
(-1)^s (q-i-2-s)^{(b-i-2-s)} \cr\cr
& & \times \biggl [ (q-b) \biggl [ {b-i-1 \choose s} (b-i-1) + {b-i
\choose s} (q-b) + (b-i-1-s) {b-i \choose s} \biggr ] \cr\cr 
& & + (b-i-1)(b-i-2) {b-i-2 \choose s} \biggr ] \cr\cr
& \ge & 0 \quad {\rm for} \ 0 \le i \le b-2 \ {\rm and} \ q \ge b \ ,
\label{theorem3c}
\eeqs
where the equal sign is realized only when $i=b-2$ at $q=b$. Therefore, we
have $\lambda ^\prime _{B_m(b),0} > \lambda ^\prime _{B_m(b),1} > ... >
\lambda ^\prime _{B_m(b),b/2}$ for even $b$ and $\lambda ^\prime 
_{B_m(b),0} > \lambda ^\prime _{B_m(b),1} > ... > \lambda ^\prime 
_{B_m(b),(b-1)/2}$ for odd $b$. Recall $f_i(b,q)=0$ if $b < i$. To show
$\lambda ^\prime _{B_m(b),b/2} > ... > \lambda ^\prime _{B_m(b),b}$
for even $b$ and $\lambda ^\prime _{B_m(b),(b+1)/2} > ... > \lambda ^\prime
_{B_m(b),b}$ for odd $b$, we must have the coefficient of $f_b(b,q)$ in
$\lambda ^\prime _{B_m(b),d}$ is smaller than the sum of the coefficients
of $f_{b-2}(b,q)$ and $f_b(b,q)$ in $\lambda ^\prime _{B_m(b),d-1}$ as 
follows,
\beqs
& & {d-1 \choose b-d-1} (b-d+1)^{(b-d-1)} + {d-1 \choose b-d+1}
(b-d+1)^{(b-d+1)} - {d \choose b-d} (b-d)^{(b-d)} \cr\cr
& = & \frac{(d-1)!(b-d+1)!}{(b-d-1)!(2d-b)!2} +  
\frac{(d-1)!(b-d+1)!}{(b-d+1)!(2d-b-2)!} - \frac{d!(b-d)!}{(b-d)!(2d-b)!}
\cr\cr
& = & \frac{(d-1)!}{(2d-b)!} \frac{(b-d+1)(b-d)}{2} +
\frac{(d-1)!}{(2d-b-2)!} - \frac{d!}{(2d-b)!} \cr\cr
& = & \frac{(d-1)!}{2(2d-b)!} \biggl [ (b-d+1)(b-d) + 2(2d-b)(2d-b-1) - 2d
\biggr ] \cr\cr
& = & \frac{(d-1)!}{2(2d-b)!} (3b^2-10bd+9d^2+3b-7d) \ ,
\label{theorem4c}
\eeqs
which is larger than 0 for even $b$ with $b/2 < d \le b$ and odd $b$ with
$(b+1)/2 < d \le b$ except for $(b,d) = (2,2)$ and $(4,3)$ where it is
0. For odd $b$ and $d=(b+1)/2$, there is no $f_b(b,q)$ term in $\lambda
^\prime _{B_m(b),(b-1)/2}$, and we need the coefficient of the $f_b(b,q)$
in $\lambda ^\prime _{B_m(b),(b+1)/2}$ is smaller than the coefficient
of $f_{b-2}(b,q)$ in $\lambda ^\prime _{B_m(b),(b-1)/2}$,
\beqs
& & {d-1 \choose b-d-1} (b-d+1)^{(b-d-1)} - {d \choose b-d} (b-d)^{(b-d)}
\cr\cr
& = & \frac{(d-1)!(b-d+1)!}{2(b-d-1)!(2d-b)!} - 
\frac{d!(b-d)!}{(b-d)!(2d-b)!} \cr\cr
& = & \frac{(d-1)!}{2(2d-b)!} \biggl [ (b-d+1)(b-d) - 2d \biggr ]  
\cr\cr
& = & \frac{1}{2} (\frac{b-1}{2})! \biggl [ 
(\frac{b+1}{2}) (\frac{b-1}{2}) - (b+1) \biggr ] \cr\cr
& = & \frac{1}{8} (\frac{b-1}{2})! (b-5)(b+1) \ge 0 \quad {\rm for} \ b
\ge 5 \ .
\label{theorem4d}
\eeqs
Therefore, for $b \ge 4$, we find $\lambda _{B_m(b),0,1}$ is dominant for
$q \ge b$, i.e. $q_c(B_m(b)) < b$.
\ $\Box$. 

We remark that our theorem, in combination with the known results
that $q_c(B_m(b))=b$ for $b=2,3$, yields the theorem that 

\beq
q_c(B_m(b)) \le b \quad {\rm for} \quad b \ge 2 \ . 
\label{theorem3following}
\eeq

We observe that the length between $b$ and $q_c(B_m(b))$ are $-1, 0, 0, 0.3264,
0.4937$, and 0.6764 for the respective values $1 \le b \le 6$. These values
suggest that this length might well increase monotonically up to 1 as $b \to
\infty$, since the lower bound for $q_c(B_m(b))$ is $b-1$ \cite{kb}.

\bigskip

We observe the following general form of four $\lambda_{B_m(b),3,j}$
corresponding to the partition $[21]$,
\beq
\lambda_{B_m(b),3,j_1} = -f_3(b,q) + 2(b-3)f_4(b,q) - (b-3)(b-4)f_5(b,q)
\label{lamj1}
\eeq
\beq
\lambda_{B_m(b),3,j_2} = -f_3(b,q) + (b-4)f_4(b,q) + (b-4)f_5(b,q)
\label{lamj2}
\eeq
\beq
\lambda_{B_m(b),3,j_3} = -f_3(b,q) + (b-6)f_4(b,q) + 3(b-4)f_5(b,q)
\label{lamj3}
\eeq
\beq
\lambda_{B_m(b),3,j_4} = -f_3(b,q) - 3f_4(b,q) - 3f_5(b,q) \ .
\label{lamj4}
\eeq
These formulae are correct for $3 \le b \le 9$, and we conjecture that
these are the only $\lambda_{B_m(b),3,j}$ corresponding to the partition
$[21]$ for arbitrarily $b$. For toroidal strips, their coefficients are 
$c_{B_m(b),3,j} = \frac{1}{6}q(q-2)(q-4)$, and $c^\prime_{B_m(b),3,j_k} =
2(b-1), b(b-3), (b-1)(b-2)$, and $\frac{2}{3}b(b-2)(b-4)$ for $1 \le k
\le 4$. 

\bigskip

From toroidal strips to Klein bottle strips, we observe the
transformations from $c^\prime_{{B}_m(b),d,j}$ to  $c^\prime_{{\bar
B}_m(b),d,j}$ have certain patterns for $0 \le d \le 4$ and $2 \le b \le
6$, and summarized them in table \ref{ctable}. Notice that the results are
different for even and odd $b$.

\newpage

\begin{table}
\caption{\footnotesize{Relation of $c^\prime_{{B}_m(b),d,j}$ and
$c^\prime_{{\bar B}_m(b),d,j}$}}
\begin{center}
\begin{tabular}[hbtp]{|c|c||c|c|c|}
partition & $d$ & $c^\prime_{{B}_m(b),d,j}$ & $c^\prime_{{\bar
B}_m(b),d,j}$ for odd $b$ & $c^\prime_{{\bar B}_m(b),d,j}$ for even $b$ \\
\hline\hline
$[d]$ & $d \ge 0$ & 1 & 1 & 1 \\
      & $d \ge 1$ & $b-1$ & 0 & $-1$ \\
      & $d \ge 2$ & $b(b-3)/2$ & $(b-3)/2$ & $b/2$ \\
      & $d \ge 3$ & $b(b-1)(b-5)/6$ & 0 & $-b/2$ \\ 
      & $d \ge 4$ & $b(b-1)(b-2)(b-7)/24$ & $(b-1)(b-7)/8$ & $b(b-2)/8$ \\
         \hline
$[1^d]$ & 2 & $b-1$ & 0 & $-1$ \\
        & 2,3 & $(b-1)(b-2)/2$ & $-(b-1)/2$ & $-(b-2)/2$ \\
        & 3,4 & $(b-1)(b-2)(b-3)/6$ & 0 & $(b-2)/2$ \\ 
        & 4,5 & $(b-1)(b-2)(b-3)(b-4)/24$ & $(b-1)(b-3)/8$ &
                $(b-2)(b-4)/8$ \\ \hline
$[21]$ & 3 & $2(b-1)$ & 0 & $-2$ \\
       & 3 & $b(b-3)$ & $b-3$ & $b$ \\
       & 3 & $(b-1)(b-2)$ & $-(b-1)$ & $-(b-2)$ \\
       & 3 & $2b(b-2)(b-4)/3$ & 2 & 0 \\ \hline
$[31]$ & 4 & $3(b-1)$ & 0 & $-3$ \\
       & 4 & $3b(b-3)/2$ & $3(b-3)/2$ & $3b/2$ \\
       & 4 & $3(b-1)(b-2)/2$ & $-3(b-1)/2$ & $-3(b-2)/2$ \\
       & 4 & $b(b-1)(b-5)/2$ & 0 & $-3b/2$ \\
       & 4 & $b(b-2)(b-4)$ & 3 & 0 \\
       & 4 & $3b(b-1)(b-3)(b-6)/8$ & $-3(b-1)(b-3)/8$ & $-3b(b-6)/8$ \\
\hline
$[22]$ & 4 & $b(b-3)$ & $b-3$ & $b$ \\
       & 4 & $2b(b-2)(b-4)/3$ & 2 & 0 \\
       & 4 & $b(b-1)(b-4)(b-5)/6$ & $(b-1)(b-5)/2$ & $b(b-4)/2$ \\ \hline
$[211]$ & 4 & $3(b-1)(b-2)/2$ & $-3(b-1)/2$ & $-3(b-2)/2$ \\
        & 4 & $(b-1)(b-2)(b-3)/2$ & 0 & $3(b-2)/2$ \\
        & 4 & $b(b-2)(b-4)$ & 3 & 0 \\
        & 4 & $3b(b-2)(b-3)(b-5)/8$ & $-3(b-3)(b-5)/8$ & $-3b(b-2)/8$ \\
\end{tabular}
\end{center}
\label{ctable}
\end{table}

For the partition $[d]$, if we label $c^\prime_{{B}_m(b),d \ge 0}$ as the
$c^\prime_{{B}_m(b),d,j} = 1$ in the first row of table \ref{ctable}, and
similarly for other $c^\prime_{{B}_m(b),d \ge d^\prime}$, we conjecture
the following forms of the corresponding $c^\prime_{{\bar B}_m(b),d \ge 
d^\prime}$ for general $d^\prime$,
\beq
c^\prime_{{\bar B}_m(b),d \ge d^\prime} = \cases{ {(b-1)/2 \choose 
d^\prime/2}-{(b-1)/2 \choose d^\prime/2-1} & for even $d^\prime \ge 2$ \cr
0 & for odd $d^\prime \ge 1$ } \quad {\rm for \ odd \ } b \ ,
\label{pcprimeoddb}
\eeq
and
\beq
c^\prime_{{\bar B}_m(b),d \ge d^\prime} = \cases{ {b/2 \choose d^\prime/2}
& for even $d^\prime \ge 0$ \cr   
-{b/2 \choose (d^\prime-1)/2} & for odd $d^\prime \ge 1$ } \quad {\rm for
\ even \ } b \ .
\label{pcprimeevenb} 
\eeq                                    

For the partition $[1^d]$, if we label $c^\prime_{{B}_m(b),2}$ as the 
$c^\prime_{{B}_m(b),d,j} = b-1$ in the first row of the $[1^d]$ partition
of table \ref{ctable}, and similarly for other 
$c^\prime_{{B}_m(b),d^\prime}$, we  conjecture the following forms of the
corresponding $c^\prime_{{\bar B}_m(b),d^\prime}$ for general $d^\prime$,
\beq
c^\prime_{{\bar B}_m(b),d^\prime} = \cases{ (-1)^{(d^\prime-1)/2}
{(b-1)/2 \choose (d^\prime-1)/2} & for odd $d^\prime \ge 3$ \cr
0 & for even $d^\prime \ge 2$ } \quad {\rm for \ odd \ } b \ ,
\label{acprimeoddb} 
\eeq
and
\beq
c^\prime_{{\bar B}_m(b),d^\prime} = \cases{ \sum ^{(d^\prime -1)/2}
_{j=0} (-1)^j {b/2 \choose j} & for odd $d^\prime \ge 3$ \cr
\sum ^{(d^\prime -2)/2} _{j=0} (-1)^{j-1} {b/2 \choose j} & for even
$d^\prime \ge 2$ } \quad {\rm for \ even \ } b \ ,
\label{acprimeevenb}
\eeq

\bigskip

If the sum of the coefficients for a specific $d$ is denoted as $C({\bar
B}_m(b),d,q)$, that is,
\beq
C({\bar B}_m(b),d,q)=\sum_{j=1}^{N_{{\bar B}_m(b),d,\lambda}}
c_{{\bar B}_m(b),d,j}(q)c^\prime_{{\bar B}_m(b),d,j} \ ,
\label{cblsum}
\eeq
we have the following conjecture,
\beq
C({\bar B}_m({\rm even}\ b),d,q)=\cases{ (-1)^{d/2}{b/2 \choose d/2} & for
                                           even $d$ \cr 
                                        0 & for odd $d$ }
\label{cevenblsum}
\eeq
\beq
C({\bar B}_m({\rm odd}\ b),d,q)=\cases{ (-1)^{d/2}{(b-1)/2 \choose d/2} &
                                          for even $d$ \cr
(-1)^{(d-1)/2} (q-1) {(b-1)/2 \choose (d-1)/2} & for odd $d$ } \ .
\label{coddblsum}
\eeq

\section{Conclusions}

In this paper we have presented exact zero-temperature partition functions for
families of chain graphs composed of repeated complete subgraphs $K_b$, $b=5,6$
with periodic or twisted periodic boundary condition in the longitudinal
direction. In the $L_x \to \infty$ limit, the continuous accumulation set of
chromatic zeros ${\cal B}$ was determined. We showed the eigenvalues with
coefficients of degree $b-1$ are bounded between $(-1)^{b-1} (q-1)$ and
$(-1)^{b-1} (q-2b+1)$, and gave the explicit forms of the eigenvalues for
partitions of the form [21]. The minimum real $q$ at which ${\cal B}$ crosses
the real $q$ axis was proven to be $q=0$, and $q_c(B_m(b)) < b$ for $4 \le
b$.

\vspace{10mm}

Acknowledgment: I would like to thank Prof. R. Shrock for helpful
discussions.

\vfill
\eject

\end{document}